\documentclass[11pt,letterpaper]{article}
\usepackage{fullpage,indentfirst,rotating}
\usepackage{float}
\usepackage{color}

\usepackage{epsfig}
\usepackage{rotating}
\usepackage{subfigure}
\usepackage{amsfonts}
\usepackage{latexsym}
\usepackage{amsmath}
\usepackage{amssymb}
\usepackage{amsthm}
\usepackage{amstext}
\usepackage[utf8]{inputenc}
\usepackage[english]{babel}
\usepackage{mathbbol}
\usepackage{bigints}
\usepackage{longtable,array}
\usepackage{longtable}
\usepackage{mathtools}
\usepackage{bbm}
\usepackage{bm}
\usepackage{booktabs}
\usepackage{setspace}
\usepackage{xcolor}
\usepackage{blindtext}
\usepackage{sectsty}
\usepackage{todonotes}
\usepackage{bigints}
\usepackage{multicol}
\usepackage{multirow}
\usepackage{comment}
\usepackage[flushleft]{threeparttable}

\sectionfont{\centering}
\subsectionfont{\centering}

\usepackage[colorlinks=true, linkcolor=blue, citecolor=blue]{hyperref}
\usepackage{soul}
\usepackage[authoryear]{natbib}
\usepackage[top=1 in, bottom=1 in, left=1 in , right=1 in]{geometry}

\allowdisplaybreaks

\newcommand{\ap}{\alpha}

\newcommand{\dt}{\delta}

\newcommand{\tht}{\theta}

\newcommand{\lt}{\left}
\newcommand{\rt}{\right}

\newcommand{\ben}{\begin{enumerate}}
\newcommand{\een}{\end{enumerate}}
\newcommand{\bit}{\begin{itemize}}
\newcommand{\eit}{\end{itemize}}

\newcommand{\eq}[1]{\begin{align}#1\end{align}}
\newcommand{\eqs}[1]{\begin{align*}#1\end{align*}}

\newcommand{\bft}{\mathbf{t}}
\newcommand{\coloneq}{:=}

\DeclareMathOperator*{\argmax}{arg max}
\DeclareMathOperator*{\argmin}{arg min}

\providecommand{\keywords}[1]{\textit{Keywords:} #1}
\providecommand{\codes}[1]{\textit{JEL Classification:} #1}

\newtheorem{assum}{Assumption}
\newtheorem{cor}{Corollary}

\newtheorem{theorem}{Theorem}
\newtheorem{lemma}{Lemma}
\newtheorem{remark}{Remark}

\theoremstyle{definition}
\newtheorem{exmp}{Example}[section]

\title{Statistical Treatment Rules under Social Interaction\thanks{
We would like to thank Kei Hirano, Simon Lee, and seminar participants at SNU, KAEA, Toronto, and Penn State for helpful comments. 
Han and Shin are grateful for partial support by the Social Sciences and Humanities Research Council of Canada.}
}

\author{
  Seungjin Han\thanks{Department of Economics, McMaster University, Email:
  \texttt{hansj@mcmaster.ca}}
\and
  Julius Owusu\thanks{Department of Economics, McMaster University, Email:
  \texttt{owusuj4@mcmaster.ca}}
\and
 Youngki Shin\thanks{Corresponding Author. Department of Economics, McMaster University, Email:
  \texttt{shiny11@mcmaster.ca} }
}

\vspace{2cm}
\begin{document}

\doublespacing

\maketitle
\begin{abstract}
\thispagestyle{empty}
In this paper we study treatment assignment rules in the presence of social interaction.
We construct an analytical framework under the anonymous interaction assumption, where the decision problem becomes choosing a treatment fraction.
We propose a multinomial empirical success (MES) rule that includes the empirical success rule of \cite{manski2004statistical} as a special case.
We investigate the non-asymptotic bounds of the expected utility based on the MES rule.
Finally, we show that the MES rule achieves the asymptotic optimality with the minimax regret criterion.
\newline
\keywords{statistical treatment rules, social interaction, finite action problems, minimax regret, optimality.}
 \newline
 \codes{C01, C44.}
\end{abstract}


\setlength{\parindent}{3ex}
\newpage

\setcounter{page}{1}

\section{Introduction}

One of the most crucial questions for a policy maker is how to assign a treatment to an individual or a group. For example, during the COVID-19 pandemic, each government has tried to find an effective order of vaccination. Recently, statistical treatment rules based on the decision theoretic framework have received much attention in treatment evaluation studies (for a general review, see \citet{manski2004statistical,manski2021econometrics} and \citet{hirano2020asymptotic}). Compared to the conventional approaches based on the point estimation and inference procedures, statistical treatment rules make it possible to evaluate a broader range of treatment rules, which includes a direct map from data to an action. Despite active research in this area, most studies focus on the individualistic treatment response and we have limited results for the case where treatment outcomes depend on each other. As we can see from the vaccination example, it is important in many empirical settings to consider dependent treatment outcomes

In this paper we study a treatment assignment rule in the presence of treatment outcome dependency.
In addition to the problem of vaccination, there are many applications that a policy maker has to weigh dependent treatment outcomes.
\citet*{heckman1999human} evaluate the effect of a tuition reduction policy in the UK in a general equilibrium framework.
They show that ignoring the outcome dependency over-estimates the effect of the policy on college enrollment more than 10 times.
\citet{duflo2004scaling} also argues that even a randomized control trial faces a challenge in scaling up to a larger level because of the general equilibrium effects or, more generally, dependent treatment outcomes. 
Using Danish data on a large job assistance program, \citet{gautier2018estimating} show that the unemployed who are not selected in the program spend more time in job search than those who look for a job in provinces without such a program.
Thus, the outcome of the untreated depends on that of the treated, and the treatment evaluations assuming independent treatment outcomes can mislead a policy maker.\footnote{See also  \cite{beaman2012social}, \cite{bursztyn2014understanding}, and \cite{duflo2003role} for additional examples.}

We investigate this problem in the framework of the statistical decision theory.
Treatment outcomes are allowed to depend on each other in a flexible way. We aim to construct a treatment assignment rule under the minimax regret approach and to characterize it.
Thus, a treatment choice using sample data, i.e.\ a statistical decision rule, is the main object of interest in this paper.
Having in mind a large-scale policy implementation, we do not impose any individual network information available. Instead, we impose a shape restriction on treatment response functions following \cite{manski2013identification}.
Specifically, we assume \emph{anonymous interactions}, which implies that the treatment response of an individual does depend on the treatment status of others but is invariant of the identity of other individuals.
In other words, it is independent of the permutation of the treatment assignments on others.
In the job assistance program above, for instance, this condition implies that the negative effect of the policy on the untreated only depends on the total size of people who receive the benefit of the job assistance program.
This assumption provides a good approximation of the world with a large-scale policy implementation, and it makes both theoretical and empirical analyses feasible by reducing the domain of the response function substantially.

We define the sampling process carefully following the statistical decision theory framework.
It contrasts to the standard \emph{individualistic} treatment effect model in that our process represents both the treatment status variable and the outcome variables as a vector. 
The dimension of the vector is the same as the number of different treatment ratios in the target population. We adopt the minimax regret approach to handle the underlying ambiguity of the data generating process.
We propose an intuitive decision rule called the multinomial empirical success (MES) rule that extends the empirical success rule in \citet{manski2004statistical} to the current setup.
We investigate the properties of the MES rule followed by the possible applications.

The main contributions of this paper are summarized as follows.
First, we prove that the MES rule achieves the asymptotic optimality for the minimax regret criterion. Using the structure of the finite action problem in statistics literature, it extends the seminal optimality result in \citet{hirano2009asymptotics} to multiple treatments.
Second, we derive the non-asymptotic bounds of the expected welfare and the maximum regret under the MES rule.
It is challenging to obtain these bounds since outcomes are correlated under social interaction.
We also provide two applications on how these bounds can be used: (i) designing an optimal sampling procedure, and (ii) computing the sufficient sample size to allow additional covariates in the treatment rule.

The rest of the paper is organized as follows.
We finish this section by reviewing related literature.
In section \ref{sec:framework} we provide the main framework of the analysis.
In section \ref{sec:multinomial} we define the MES rule and derive the upper bounds of the maximum regret.
We also provide two applications of these bounds.
In section \ref{sec:optimality} we show the asymptotic optimality of the MES rule.
We provide some concluding remarks in section \ref{sec:conclusion}. All proofs and technical details are deferred to the appendix.

\subsection{Related Literature}
In the seminal work of \cite{manski2004statistical}, he considers the statistical decision theory in the context of heterogeneous treatment rules. He proposes the empirical success rule and derives the finite sample bounds of the minimax regret.
\citet{stoye2009minimax} characterizes the minimax regret rule using the game theoretic approach and shows that the empirical success rule is a good approximation of the minimax regret rule under certain sampling processes.
\citet{hirano2009asymptotics} apply the limit experiment framework to develop large sample approximations to the statistical treatment rules.

\citet{kitagawa2018should} propose the empirical welfare maximization (EWM) method that selects the treatment rule maximizing the sample analogue of the social average welfare.
\citet{athey2021policy} propose a doubly robust estimation procedure for the EWM problem and show the rate-optimal regret bounds.
\cite{mbakop2021model} consider a large class of admissible rules and propose a penalized EWM method that chooses the optimal size of the policy class.
\citet{manski2016sufficient,manski2019trial} argue to design clinical trials based on the goal of statistical treatment rules rather than on the statistical power of a hypothesis test.
Motivated by a risk-averse policy maker, \citet{manski2007admissible} and \citet*{kitagawa2022treatment} propose nonlinear transformations of welfare and regret.

\citet{manski2013identification} studies identification of treatment effects with social interaction. To make the problem feasible, he proposes possible approximation methods including \emph{anonymous interaction}, which will be explained in detail later. \citet{manski2009identification} analyzes statistical treatment rules under the anonymous interaction assumption and the shape restriction on the mean welfare function. \citet{viviano2019policy} proposes the network empirical welfare maximization method under the anonymous interaction assumption among those in the first-degree neighbor. However, our approach is different from his since it does not require heavy computation to solve an empirical optimization problem. It is also new that the proposed multinomial empirical success rule achieves the asymptotic optimality in the sense of \citet{hirano2009asymptotics}.

\section{Framework}\label{sec:framework}
We consider the following framework based on \cite{manski2004statistical} and  \cite{stoye2009minimax}.
Consider a social planner who assigns a binary treatment  $T \in \{0,1\}$ to each individual $j$ in a heterogeneous population  $J$.
The population is divided into mutually exclusive and exhaustive groups based on observed characteristics (e.g.\ high school graduate vs. college graduate).
Let $g \in \{1,2,\ldots,G\}$ be the index of a group and $n_{g}$ be the (population) size of group $g$.
Individual $j$  in group $g$ has a response function $y_{jg} :\{0,1\}\times\{0,1\}^{n_{g} -1} \mapsto  [0,1]$ that maps each possible group treatment vector $\mathbf{t} = (t_1,\dots,t_{n_{g}}) \in \{0,1\}^{n_{g}}$ into an outcome in $[0,1]$.
Thus, we can write $ y_{jg}(\mathbf{t}) = y_{jg}(t_j, \mathbf{t}_{-j}) $, where $t_j$  is the treatment assigned to individual $j$  and   $\mathbf{t}_{-j}$  represents the treatment vector for individuals in the same group  excluding person $j$'s treatment assignment.
This response function generalizes the individualistic treatment  in a way that the spillover effect is allowed inside the same group (e.g.\ segmented labour markets).
Note that the model allows the most flexible interactions when the whole population is categorized as a single group.
The range of $[0,1]$ is a simple normalization and any bounded outcome space can be allowed.
For notational simplicity, we consider a single group from now on and drop the subscript $g$ unless it causes any confusion.


We consider a probability space  $(J, \Sigma,  P_J)$. The population $J$ is dense in the sense that  $ P_J(\{j\})=0$,  for all   $j\in J$.
The social planner cannot distinguish members of $J$. Therefore, we can consider the model as an induced random process, $Y(\mathbf{t})$, 
which is a potential outcome depending not only on individual treatment status, $t_j$, but on possible treatments of other members, $\mathbf{t}_{-j}$. Given the large size of the population $J$, this random process in the most general structure is intractable. Following the social interaction literature, we impose the following assumption.

\begin{assum}[Anonymous Interactions, \cite{manski2013identification}]\label{assump:anonymous}
  The outcome of individual $j$ is invariant with respect to permutations  of the treatments received by other members of the group.
\end{assum}
Assumption \ref{assump:anonymous} implies that a treatment ratio is a sufficient statistic for $\mathbf{t}_{-j}$. Let $\pi(\mathbf{t})$ be a treatment ratio of treatment vector $\mathbf{t}$. Then, for two treatment vectors $\bft \neq \bft'$ such that $\bft= (t_j, \bft_{-j})$ and $\bft'= (t_j, \bft'_{-j})$, Assumption \ref{assump:anonymous} implies that
\eqs{
 y_j(\bft)=y_j(\bft') \mbox{ if }\pi(\bft)=\pi(\bft').
 }
 Therefore, the outcome of a treatment $\bft$ depends on individual's treatment status $t_j$ and  $\pi(\mathbf{t})$, and we can rewrite the the response function $y_j(\bft)$ as $y_j(t_j,\pi(\bft_{-j})): \{0,1\} \times \Pi \mapsto [0,1]$, where $\Pi:=[0,1]$. The potential outcome processes now become $(Y_0(\pi),Y_1(\pi))$ whose distribution is $P_Y(Y_0(\pi),Y_1(\pi))$. Note that the induced measure $P_Y$ can be constructed from $P_J$ given the response function $y_j(\cdot)$.

The distribution $P_Y$ is identified with a state of the world $\theta \in \Theta$ that is unknown to the policy maker. Note that $\{P_{Y,\theta}(Y_0(\pi),Y_1(\pi)): \theta \in \Theta\}$ is composed of all possible distributions on the outcome space $[0,1]^2$ for each $\pi \in \Pi$.
To make the main arguments clear, we impose an additional assumption that the set $\Pi$ is discrete.
\begin{assum}[Discrete Choice Set]\label{assump:discrete Pi}
  Let  $\pi$  be the fraction of  treated individuals in a group. The  support  of  $\pi$  denoted by  $\mathbf{\Pi}$ is a discrete set of finite elements.
\end{assum}
Assumption \ref{assump:discrete Pi} is suitable to many applied settings since the treatment ratio set may be  constrained exogenously  for ethical, budgetary, equity, legislative  or political reasons.    In addition, this is a practical assumption when experiments are costly to implement at all feasible treatment ratios. The assumption could also provide a good approximation if $\mathbf{\Pi}$ is a continuous interval but outcome function $y_j$ is smooth in $\pi$

We provide the following examples below.
\begin{exmp}[Job placement assistantship program] \label{Job placement assistantship program} \citet{crepon2013labor} design a two-stage randomized experiment to evaluate the direct and displacement impacts of job placement assistance (JPA) on the labor market outcomes of young, educated job seekers in France. Individuals are organized in  segmented labour markets (e.g. cities) and five treatment ratios (0\%, 25\%, 50\%, 75\%, and 100\%) are considered. An individual's labor market outcome depends not only on his/her treatment status but on the treatment ratio (fraction of individuals who received the JPA in their labor market).
\end{exmp}

\begin{exmp} [Cholera vaccine coverage] \cite{root2011role} analyze data from a field trial  in  Bangladesh  to  assess the  evidence  of  indirect  protection  from  cholera  vaccines  when  vaccination coverage rates varies according to the social  network. Households are organized into  independent groups using kinship connections. Vaccine  coverage rate is discretized into the following ranges: $(0,27.2\%]$, $(27.2,40.0\%]$, $(40.0-50.0\%]$, $(50.0\%-62.5\%]$, and $(62.5\%, 100\%]$.

\end{exmp}

We now turn our attention to a random sample that helps the policy maker infer the state of the world $\theta$.
Let $\mathbf{\Pi}= \{\pi_1, \pi_2,\dots, \pi_K\}$.
The experiment generates a sample space   $\Omega := ( \{0,1\} \times [0, 1])^{n}$, where $n :=\sum_{k=1}^K n_{k}$ and $n_{k}$ is the subgroup size of an experiment with a treatment ratio $\pi_k$.
A typical element of $\Omega$ is represented by
\eqs{
\omega^n := \{(t_{i_1}(\pi_1) ,y_{i_1}(\pi_1))_{i_1=1}^{n_{1}},(t_{i_2}(\pi_2) ,y_{i_2}(\pi_2))_{i_2=1}^{n_{2}}, \dots , (t_{i_K}(\pi_K) ,y_{i_K}(\pi_K))_{i_K=1}^{n_{K}} \}.
}
Conditional on the treatment $t_{n_k}(\pi_k)$,  $y_{n_k}(\pi_k)$ is an independent realization of $Y_{t}(\pi_k)$ for $t=0,1$. Therefore, it helps a policy maker to infer the state of the world $\theta$. To make notation simple, we assume the equal subgroup size, $n_1=\cdots=n_K = n/K$, and $\omega^n$ is composed with n-copies of
$
\omega_i := \{(t_{i}(\pi_1) ,y_{i}(\pi_1)), \dots , (t_{i}(\pi_K) ,y_{i}(\pi_K)) \}.
$

The policy maker constructs a statistical treatment rule  $\delta : \Omega \mapsto \mathbf{\Pi}$\, that  maps a sample realization  $\omega^n$  onto a treatment assignment ratio  $\pi \in \Pi$.
Recall that we restrict our attention to a single group in this framework but the statistical treatment rule can be group-specific when there are multiple groups. In  section \ref{sec:extensions}, we  extend the current frame to the multiple groups case.

The expected outcome (or social welfare) given the statistical treatment rule $\delta$ and the state $\theta$ is
\eq{
u(\dt,\theta)  & := \int U(\delta(\omega^n),\theta) dQ^n_{\theta} \label{eq:expected-outcome-general} \\
          & = \sum_{k=1}^K  U(\pi_k,\theta) \Pr(\dt(\omega^n)=\pi_k ; \theta), \label{eq:expected-outcome}
}
where $Q_{\theta}$  is a distribution of $\omega_i$ given state $\theta$, $U(\pi,\theta):= (1-\pi) \cdot E_{\theta}[Y_0(\pi)] + \pi \cdot E_{\theta}[Y_1(\pi)]$ is the expected outcome (or social welfare) for any given treatment ratio $\pi$ in state $\theta$, and $E_{\theta}[Y_t(\pi)]$ is the mean potential outcome of treatment status $t$ given $\theta$ and $\pi$.
Note that the potential outcome variable $Y_t(\pi)$ depends on the treatment of others through $\pi$. This point becomes clearer if we compare the expected outcome in \eqref{eq:expected-outcome-general} with that of the individualistic treatment model (e.g.\ \citet{stoye2009minimax}). When there is no social interaction, the mean potential outcome is independent of the group treatment ratio $\pi$, i.e.\ $E_{\theta}[Y_t(\pi)]=E_{\theta}[Y_t]$. Then, the expected outcome in \eqref{eq:expected-outcome-general} becomes
\eqs{
\int U(\delta(\omega^n),{\theta}) dQ^n_{\theta}
    & = \int \left\{ (1-\delta(\omega^n)) E_{\theta}[Y_0] + \delta(\omega^n) E_{\theta}[Y_1] \right\} dQ^n_{\theta} \\
    & = E_{\theta}[Y_0] \left( 1 - \int \delta(\omega^n) dQ^n_{\theta}\right) + E_{\theta}[Y_1] \int \delta(\omega^n) dQ^n_{\theta} \\
    & \equiv \mu_0 (1- E_{\theta}[\dt(\omega)]) + \mu_1 E_{\theta}[\dt(\omega)],
}where the last line is equal to the expected outcome in \citet{stoye2009minimax} using his notation.

It is interesting to compare our framework to the individualistic multiple-treatment design. Given the finite number of treatment ratios, one might want to interpret the framework in terms of $K$ different individual treatments without any social interaction: e.g.\ define $Y_1 := Y_1(\pi_1), Y_2 := Y_1(\pi_2), \ldots, Y_K:= Y_1(\pi_K)$ and set $(Y_0,Y_1,\ldots,Y_K)$ as a vector of potential outcomes. However, this multiple-treatment design does not capture the feedback effect of the social interaction for any non-treated individual. Note that $Y_0(\pi)$ still depends on the treatment ratio $\pi$ in our framework, which is not embedded in the potential outcome vector $(Y_0,Y_1,\ldots,Y_K)$ of the standard multiple-treatment design.

The decision problem is to find a statistical treatment rule that maximizes the expected outcome function $u(\delta,\theta)$. However, there exists ambiguity in the sampling process and we need some decision criteria for unknown $\theta$. In this paper we adopt the minimax regret rule following \citet{manski2004statistical} and \citet{stoye2009minimax}. The regret function of $\delta$ given state $\theta$ is defined as
\eq{
R(\dt,{\theta}) := \max_{d \in D} u(d,{\theta}) - u(\dt,{\theta}), \label{eq:regret}
}where $D$ is a set of all possible statistical treatment rule. The minimax regret solution of the decision problem is defined as
\eq{
  \dt^* := \argmin_{\dt \in D} \sup_{{\theta} \in \Theta} R(\dt,{\theta}). \label{eq:optimal-solution}
}

\section{Multinomial Empirical Success Rule}\label{sec:multinomial}
In this section we propose a feasible statistical decision rule and characterize it by the non-asymptotic bounds on the maximum regret.
To show the main idea, we keep focusing on a single group case.
The results are extended into the multiple-group case in section \ref{sec:extensions} and we show how they can be used to determine the proper level of groups.

It is difficult to attain the optimal statistical treatment rule by solving \eqref{eq:optimal-solution} directly since $R(\delta,\theta)$ involves integration over finite sample distributions. As an alternative, researchers may propose possible statistical treatment rules and analyze whether they achieve the optimal regret level. One of the popular rules is an empirical success rule, which substitutes empirical success rates for the population counterparts.

We propose such an empirical success rule suitable for the proposed setup. To focus on our main arguments, we restrict our attention to samples with a strict ordering of the  estimates for ${U}(\pi,s)$  for all  $\pi \in \mathbf{\Pi}$.
We define our multinomial empirical success (MES) rule as follows:
\begin{align}
      \delta^{MES}(\omega) \coloneq  \sum_{k=1}^K \pi_k \cdot   \mathbbm{1}\left(\hat{U}(\pi_k)>\max_{\pi \in \mathbf{\Pi}_{-k} } \hat{U}(\pi)\right), \label{eq:MES}
\end{align}
where $\mathbf{\Pi}_{-k} := \mathbf{\Pi} \setminus \{\pi_k\}$ and
\begin{align}
\hat{U}(\pi_k)
  & \coloneq (1 - \pi_k) \cdot \hat{E}_{P_{\theta}}[Y_0(\pi_k)] + \pi_k \cdot \hat{E}_{P_{\theta}}[Y_1(\pi_k)]  \nonumber \\
    & = (1-\pi_k)\cdot \frac{\sum_{n_k=1}^{N_{k}} Y_{n_k}(\pi_k)\cdot  \mathbbm{1}(T_{n_k}(\pi_k)=0)}{\sum_{n_k=1}^{N_{k}}   \mathbbm{1}(T_{n_k}(\pi_k)=0)}   +  \pi_k \cdot  \frac{\sum_{n_k=1}^{N_{k}} Y_{n_k}(\pi_k)\cdot  \mathbbm{1}(T_{n_k}(\pi_k)=1)}{ \sum_{n_k=1}^{N_{k}}  \mathbbm{1}(T_{n_k}(\pi_k)=1)}. \label{eq:empirical_U_hat}
\end{align}
Note that, using the convention $0\cdot \infty =0$, we define $\hat{U}(0)=N_1^{-1} \sum_{n_1=1}^{N_1} Y_{n_1}(0)$ when $\pi_1=0$. Similarly, $\hat{U}(1)=N_K^{-1} \sum_{n_K=1}^{N_K} Y_{n_1}(1)$ when $\pi_K=1$.

We have a few remarks on the proposed statistical decision rule.
First, we call the rule in \eqref{eq:MES} as a Multinomial Empirical Success (MES) rule to emphasize the multinomial choice set in the setting.
Second, we estimate $E_{P_{\theta}}[Y_t(\pi)]$ by using the empirical measure that depends on the unknown state $\theta$ of the world.
Thus, both $\hat{U}(\pi_k)$ and the outcome of $\delta^{MES}(\cdot)$ depend on $\theta$ although it is not included as an argument explicitly.
Third, the MES rule encompasses the (unconditional) empirical success rule in \citet{manski2004statistical}.
Let $\Pi=\{0,1\}$ with $\pi_1=0$ and $\pi_2=1$.
Then, the MES rule becomes
\begin{align*}
    \delta^{MES}(\omega) & = 0 \cdot   \mathbbm{1}\lt(\hat{U}(0,{\theta})>\hat{U}(1,{\theta})\rt) + 1 \cdot   \mathbbm{1}\lt(\hat{U}(0,{\theta})< \hat{U}(1,{\theta})\rt)\\
                         & =  \mathbbm{1}\lt(  \frac{1}{N_1} \sum_{n_1=1}^{N_1} Y_{n_1}(0) < \frac{1}{N_2} \sum_{n_2=1}^{N_2} Y_{n_1}(1) \rt),
\end{align*}
which is the empirical success rule in \citet{manski2004statistical}.

We next evaluate the expected outcome in \eqref{eq:expected-outcome} using the MES rule in \eqref{eq:MES}:
\begin{align*}
u(\delta^{MES},\theta)
    &= \sum_{k=1}^K  \Pr(\delta^{MES}(\omega^n)=\pi_k) \cdot U(\pi_k, {\theta})   \\
  &= \sum_{k=1}^K \Pr\lt(\hat{U}(\pi_k)>\max_{\pi \in \mathbf{\Pi}_{-k} } \hat{U}(\pi)\rt)\cdot U(\pi_k, {\theta})\\
    &= \sum_{k=1}^K \Pr\lt(\bigcap^K_{j=1, j\neq k}\{\hat{U}(\pi_k)> \hat{U}(\pi_j) \}\rt)\cdot U(\pi_k, {\theta}).
\end{align*}
As we discussed above, $u(\delta^{MES}, {\theta})$ is intractable since it involves all possible finite sample distributions.
However, building on \citet{manski2004statistical}, we can construct bounds for the expected outcome with the MES rule as follows.
\begin{theorem}\label{thm:bounds}
 Fix $\theta\in\Theta$. Let $\mathbf{\Pi} = \{\pi_1, \dots, \pi_K \}$, $\Delta_{kl} \coloneq|U(\pi_k, {\theta})- U(\pi_l, {\theta})|$ for $k,l=1,\ldots,K$, and $\pi_{M^*}:=\argmax_{\pi \in \mathbf{\Pi}}  U(\pi,{\theta})$.
 Then, the following inequality holds:
 \begin{align}
      U(\pi_{M^*}, {\theta})- \sum_{k=1}^{K} \exp \Bigg[-2\Delta_{{M^*}k}^2 \{A_{k}+A_{M^*} \}^{-1} \Bigg]\cdot \Delta_{{M^*}k}
\leq u(\delta^{MES}, {\theta})
\leq U(\pi_{M^*}, {\theta}), \label{eq:main-theorem}
\end{align}
where  $A_{k}\coloneq (1-\pi_k)^{2}N_{k0}^{-1} + \pi_{k}^{2}N_{k1}^{-1}$, $A_{{M^*}}\coloneq(1-\pi_{M^*})^{2}N_{{M^*} 0}^{-1} +  \pi_{M^*}^{2}N_{{M^*} 1}^{-1}$, and $N_{kt}$ denotes the number of individuals in the sample with $\pi=\pi_{k}$ and $T=t$.
\end{theorem}
It is worth comparing these bounds with those in Proposition 1 of \citet{manski2004statistical}.
Note that both frameworks allow the potential outcome distributions to vary across some indexing variables.
For example, the potential outcomes in \citet{manski2004statistical} depend on exogenous conditioning variables $X$, i.e.\ heterogeneous treatment effects over $X$.
However, we focus on the dependence of the potential outcomes on the choice variable $\pi$.
They look similar from the mathematical perspective, but the implications are quite different since the result in this paper allows the effect of social interaction.
This point becomes clearer when we extend the model to the case that includes additional conditioning variables.

We further investigate the finite sample penalty of the lower bound in \eqref{eq:main-theorem}, which measures the possible difference of $u(\delta^{MES}, \theta)$ from the ideal solution $U(\pi_{\pi_{M^*}}, \theta)$.
First, the penalty converges to zero at the exponential rate as $N_{tk}$ increases uniformly for all $t\in \{0,1\}$ and $k \in \{1, \dots, K\}$.
Second, the penalty is maximized when $\Delta_{M^*k}=\{A_{k}+A_{M^*}\} ^{1/2}/2$ for each $k \neq M^*$.
Thus, we can compute the upper bound of the penalty as follows:
 \begin{align}
     \sum_{k=1}^{K} \exp \Bigg[-2\Delta_{{M^*}k}^2 \{A_{k}+A_{{M^*}}\}^{-1} \Bigg] \cdot \Delta_{{M^*}k}
     \leq
     \frac{1}{2} \cdot e^{-\frac{1}{2}}\sum_{k=1, k \neq M^*}^{K} \{A_{k}+A_{{M^*}}\}^{\frac{1}{2}}. \label{eq:ub-of-penalty}
 \end{align}
Third, it is interesting to investigate the relationship between the cardinality of $\Pi$ denoted by $K$ and the penalty size. Consider the following example of two possible choice sets $\Pi_1$ and $\Pi_2$ such that $\Pi_1 \subset \Pi_2$.
Let $\pi_{M^*}$ be the optimal solution of $\Pi_1$.
If $\pi_{M^*}$ is also the optimal solution of $\Pi_2$, then $\Pi_2$ has a larger penalty than $\Pi_1$.
However, if the optimal solution of $\Pi_2$ denoted by $\pi_{M^{**}}$ is different from $\pi_{M^{*}}$, then $\Pi_2$ may have a smaller penalty than $\Pi_1$.
Note that $\Delta_{M^{**}k} > \Delta_{M^{*}k}$ for all $k \in \Pi_1$ and that there may exits some $k \in \Pi_1$ such that $\exp [-2\Delta_{{M^{**}}k}^2 \{A_{k}+A_{{M^{**}}} \}^{-1} ] <  \exp [-2\Delta_{{M^*}k}^2 \{A_{k}+A_{{M^*}} \}^{-1} ]$.
Therefore, a larger choice set may improve the finite sample lower bound only if it contains a better welfare outcome.
Finally, we investigate the uniform bound of the regret function over $\theta$. The upper bounds of the regret function with $\delta^{MES}$ is represented in terms of the penalty:
\eqs{
0
\leq
    \sup_{\theta \in \Theta} R(\delta^{MES}, \theta) 
& \leq 
     \sup_{\theta \in \Theta} \lt\{ \sum_{k=1}^{K} \exp \Bigg[-2\Delta_{{M^*}k}^2 \{A_k + A_{M^*} \}^{-1} \Bigg] \cdot \Delta_{{M^*}k} \rt\}\\
& \leq
    \sup_{\theta \in \Theta} \lt\{ \frac{1}{2} \cdot e^{-\frac{1}{2}}\sum_{k=1, k \neq M^*}^{K} \{A_{k}+A_{{M^*}}\}^{\frac{1}{2}} \rt\}.
}
Different from the result in \citet{manski2004statistical}, $A_{{M^*}}$ in the right hand side depends on $\theta$ since $\pi_{M^*}$ is defined in terms of $U(\pi,{\theta})$.
Therefore, we need an additional step to achieve the uniform bound.
Let $\overline{A} \coloneq \max_{k \in \{1,\dots, K\}}A_k$.
Note that $\overline{A} \ge A_{M^*}$ and that $\overline{A}$ is independent of $\theta$.
Then, the desired uniform bound is achieved as follows:
\begin{align}
0 \leq \sup_{{\theta} \in \Theta}R(\delta^{MES}, {\theta})\leq  \frac{1}{2} \cdot e^{-\frac{1}{2}}\sum_{k=1, A_k \neq \overline{A}}^{K} \{A_{k}+\overline{A}\}^{\frac{1}{2}}.\label{eq:upper_bound_regret}
\end{align}

These finite sample bounds give us two useful applications. First, we apply this bound to solve the quasi-optimal experiment design problem. Second, we can extend the bound to the covariate dependent treatment rule and determine the minimum sample size to adopt a finer covariate set as in \citet{manski2004statistical}. We provide these applications in the following two subsections.

\subsection{Application 1: Quasi-optimal Experiment Design}
We study the optimal experiment design problem under interference using the upper bound of the maximum regret.
Specifically, we focus on the randomized saturation design which is composed of two-stage randomized experiments (for example, see \cite{baird2018optimal}). Suppose that we are given many clusters. In the first stage, we assign different treatment ratios in $\Pi$ to each cluster \emph{randomly} according to a probability distribution $f$. In the second stage, a binary treatment is  assigned to each member of a cluster according to a treatment ratio assigned in the previous stage. Therefore, the randomized saturation design is fully characterized by a pair $(\Pi,f)$ and it encompasses other designs like clustered, block, and partial population designs commonly employed under interference.

We now consider an experiment design problem that minimizes the maximum regret. We cannot compute the exact regret function because of the ambiguity in $\theta$. Instead, we reformulate the problem as minimizing the feasible upper bound of the regret in \eqref{eq:upper_bound_regret}.

Recall that $N$ denotes the total sample size over all clusters and $\Pi=\{\pi_1, \pi_2, \cdots, \pi_K\}$ be a finite set of treatment ratios.
Since $\Pi$ is a finite set, we can write $f=\{(\alpha_1, \alpha_2, \cdots \alpha_K): \sum_{k=1}^K \ap_k = 1\}$, where $\ap_k$ is a probability mass of assigning $\pi_k$.
The subsample sizes can be written in terms of the treatment ratios and their corresponding probabilities:
 $N_{k0}= (1-\pi_k)\alpha_k N$ and $N_{k1}= \pi_k\alpha_kN$ for all $k= 1, 2,\ldots, K.$ Then, for each $A_k$, we have
 \eqs{
    A_k & = \frac{(1-\pi_k)^2}{N_{k0}} + \frac{\pi_k^2}{N_{k1}} \\
        & = \frac{(1-\pi_k)}{\ap_k N} + \frac{\pi_k}{\ap_k N} \\
        & = \frac{1}{\ap_k N},
 }which makes the optimization problem simple. Without loss of generality, let $\overline{A}=A_1$. We substitute $A_k$ in \eqref{eq:upper_bound_regret} and drop all irrelevant variables to get
 \eqs{
 &\hskip-60pt \min_{\{\ap_k\}_{k=1}^K} \sum_{k=2}^K \lt(\frac{1}{\ap_1N} + \frac{1}{\ap_k N}\rt)^{1/2} \\
 \mbox{subject to }~~ &\sum_{k=1}^K \ap_k = 1 \\
 & \ap_1 \le \ap_k \mbox{ for } k=2,\ldots,K.
 }
 Solving this optimization problem, we derive the quasi-optimal design of equal $\ap_k^*$ ($\ap^*_k=1/K$) only when $K=2$.
 It is worthwhile to note that \citet{baird2018optimal} derive the optimal randomized saturation design based on the statistical power but we focus on the maximum regret directly (see \cite{ manski2016sufficient} for further discussion).

\subsection{Application 2: Covariate-dependent Treatment Rules}\label{sec:extensions}
In this section, we extend the model and consider covariate-dependent treatment rules.
We first introduce new notation.
Let $X$ be a vector of covariates.
In the similar spirit of Assumption \ref{assump:discrete Pi}, we restrict our attention to discrete and finite covariates.
Then, we can vectorize the possible outcomes of $X$ and partition the population into $L$ different subsets denoted by $\mathcal{X}:=\{x_1,\ldots,x_L\}$.
To make notation simple, we assume a common domain of treatment ratios $\Pi$ for each $x_l$\footnote{We can allow different assignment ratio sets at the cost of extra notation, e.g.\ $\Pi:=\cup_{l=1}^L \Pi_l$, where $\Pi_l:=\{\pi_1,\ldots,\pi_{K_l}\}$ is the set of assignment ratios for $x_l$.}.
We define a statistical treatment rule as $\dt(x,\omega^n):\mathcal{X}\times \Omega \mapsto \Pi$.
Let $\boldsymbol{\pi}:=(\pi_1,\ldots,\pi_L)'$ be a vector of treatment assignment ratios, where $\pi_l$ is applied to  subgroup $x_l \in \mathcal{X}$.
Let $\boldsymbol{p}$ be a vector of population subgroup proportions.
Then, $\bar{\pi}:= \boldsymbol{p}'\boldsymbol{\pi}$ becomes the unconditional treatment ratio. Under Assumption \ref{assump:anonymous}, the response function can be rewritten as $y_j(t_j, \bar{\pi})$.

Given $\boldsymbol{\pi}$ and $\theta$, the  outcome of the subgroup with covariate $x_l$ is
\begin{align}
U_l(\boldsymbol{\pi},\theta) := (1-\pi_l) \cdot E_{\theta}\lt[ Y_0(\bar{\pi}) \vert X=x_l \rt] + \pi_l \cdot E_{\theta} \lt[Y_1 (\bar{\pi}) \vert X=x_l \rt].
\end{align}
Note that $U_l$ is affected by the treatment ratios of other covariate types through $\bar{\pi}$ as well as its own ratio $\pi_l$.
Let $\boldsymbol{\delta}(\omega^n):=(\delta(x_1,\omega^n), \ldots, \delta(x_L,\omega^n) )$ be a vector of statistical treatment rules over $\mathcal{X}$ when sample $\omega^n$ is realized, i.e.\ $\boldsymbol{\delta}(\omega^n): \Omega \mapsto \Pi^L$.
The expected outcome of the whole population is defined by the weighted sum of $U_l$:
\begin{align}
u(\boldsymbol{\delta},\theta) := \sum_{l=1}^L \lt[\int U_l(\boldsymbol{\delta}(\omega^n),\theta) dQ_{\theta }^n\rt] \Pr(X  = x_l).
\end{align}
If $\Pr(X=x_l;\theta)=1$ for some $l$, then $\pi_l=\bar{\pi}\equiv \pi$, $L=1$ and $u(\delta,\theta)=\int U(\delta(\omega^n),\theta)dQ_{\theta}^n$.
Therefore, the expected outcome becomes equation \eqref{eq:expected-outcome-general}, where there exists a single type of population.

Similar to \eqref{eq:optimal-solution}, we can define the minimax regret solution of the decision problem as
\eqs{
  \boldsymbol{\delta}^* := \argmin_{\boldsymbol{\delta} \in \boldsymbol{D}} \sup_{\theta \in \Theta} R(\boldsymbol{\delta}, \theta),
}where $R(\boldsymbol{\delta}, \theta):= \max_{ \boldsymbol{d} \in \boldsymbol{D}} u(\boldsymbol{d},\theta) - u(\boldsymbol{\delta},\theta)$ is a regret function. 
Since the expected welfare with covariate $x_l$ is affected by the treatment assignment ratios of other covariates $x_m\neq x_l$, we need to find the decision rule simultaneously over all elements in $\mathcal{X}$, i.e.\ the decision rule vector $\boldsymbol{\delta}$.
It is worth noting that, when we consider $x_l$ as a single group, this extension can be interpreted as multiple groups with interaction between groups via $\bar{\pi}$.

We now construct the multinomial empirical success rule conditional on covariate $x_l$. Note that $\Pi^L$ contains at most $K^L$ elements, $\vert \Pi^L \vert = K^L < \infty$.
Let $\boldsymbol{\pi}_k$ be a generic element of $\Pi^L$.
Then, the population (unconditional) treatment ratio is $\bar{\pi}_k=\boldsymbol{p}'\boldsymbol{\pi}_k$.
The empirical mean of $Y_t(\bar{\pi}_k)$ conditional on $x_l$ is
\eqs{
  \hat{E}_{\theta}[Y_t(\bar{\pi}_{k})|X=x_l]\coloneq\frac{\sum_{n_k=1}^{N_{k}} Y_{n_k}(\bar{\pi}_k)\cdot \mathbbm{1}(T_{n_k}(\bar{\pi}_k)=t, X=x_l)}{\sum_{n_k=1}^{N_{k}}  \mathbbm{1}(T_{n_k}(\bar{\pi}_k)=t, X=x_l)}~~\mbox{for}~~k=1,\ldots,K^L~\mbox{and}~t=0,1.
}
Finally, the conditional multinomial empirical success rule (CMES) is defined as follows:
\begin{align}
      \boldsymbol{\delta}^{CMES}(\omega^n) \coloneq  \sum_{k=1}^{K^L} \boldsymbol{\pi}_k \cdot  \mathbbm{1}[\hat{U}(\boldsymbol{\pi}_k)>\max_{\boldsymbol{\pi} \in \Pi^L_{-k} } \hat{U}(\boldsymbol{\pi})] \label{eq:CMES}
\end{align}
where $\Pi^L_{-k} := \Pi^L \setminus \{\boldsymbol{\pi}_k\}$ and
\begin{align*}
\hat{U}(\boldsymbol{\pi}_k)
  & \coloneq \sum_{l=1}^L \Pr(X=x_l)\cdot \hat{U}_l(\boldsymbol{\pi}_k) \\
  & = \sum_{l=1}^L \Pr(X=x_l)\Bigg[(1 - \pi_{kl}) \cdot \hat{E}_{\theta}[Y_0(\bar{\pi}_{k})|X=x_l] + \pi_{kl} \cdot \hat{E}_{\theta}[Y_1(\bar{\pi}_k)|X=x_l] \Bigg],
\end{align*}
where $\pi_{kl}$ is the $l$-th element of the $L$-dimensional vector $\boldsymbol{\pi}_k$.
The CMES rule $\delta^{CMES}(\omega)$ in \eqref{eq:CMES} looks similar to the (unconditional) MES rule in Section \ref{sec:multinomial}.
However, $\boldsymbol{\pi}_k$ is now an $L$-dimensional vector and the rule itself is an $L$-dimensional vector-valued function.
Let $U(\boldsymbol{\pi}_k, \theta)$ be the population counterpart of $\hat{U}(\boldsymbol{\pi})$ by replacing $\hat{E}_{\theta}$ with $E_{\theta}$.
Then, we can define the expected outcome given the CMES rule $\boldsymbol{\delta}^{CMES}$ as follows:
\begin{align*}
u(\boldsymbol{\delta}^{CMES}, \theta)
    &= \sum_{k=1}^{K^L} \Pr(\boldsymbol{\delta}^{CMES}(\omega^n)=\boldsymbol{\pi}_k; \theta) \cdot U(\boldsymbol{\pi}_k, \theta)\\
    & = \sum_{k=1}^{K^L} \Pr\lt(\bigcap^{K^L}_{j=1, j\neq k}  \{\hat{U}(\boldsymbol{\pi}_k)> \hat{U}(\boldsymbol{\pi}_j) \}; \theta \rt)\cdot U(\boldsymbol{\pi}_k, \theta).
\end{align*}

 We are now ready to extend the the bounds of the expected outcome in (\ref{eq:main-theorem}) to the CMES rule.
 \begin{theorem}\label{thm:bounds-cmes}
 Fix $\theta\in\Theta$. Let $\Pi^L = \{\boldsymbol{\pi}_1, \dots, \boldsymbol{\pi}_{K^L} \}$, $\Delta_{kj} \coloneq|U(\boldsymbol{\pi}_k, \theta)- U(\boldsymbol{\pi}_{j}, \theta)|$ for $k,j=1\ldots,K^L$, and $\boldsymbol{\pi}_{M^*}:=\argmax_{\boldsymbol{\pi} \in \Pi^L}  U(\boldsymbol{\pi},\theta)$.
 Then, the following inequality holds:
 \begin{align*}
    U(\boldsymbol{\pi}_{M^*}, \theta)- \sum_{k=1}^{K^{L}} \exp & \lt( -2\Delta_{M^{*}k}^2\cdot\lt\{\sum_{l=1}^L \Pr(X=x_l)^2 (A_{kl}+A_{M^{*}l})\rt\}^{-1} \rt)\cdot  \Delta_{M^{*}k} \\
    & \hskip140pt \leq u(\boldsymbol{\delta}^{CMES}, \theta)\leq  U(\boldsymbol{\pi}_{M^*}, \theta), \label{eq:main-corollary}
\end{align*}
where  $A_{kl}\coloneq (1-\pi_{kl})^{2}N_{k0l}^{-1} + \pi_{kl}^{2}N_{k1l}^{-1} $ and $A_{{M^*}l}\coloneq(1-\boldsymbol{\pi}_{M^* l})^{2}N_{M^*0l}^{-1} +  \pi_{l1}^{2}N_{M^*1l}^{-1},$ with $N_{ktl}$ representing the number of individuals with $\boldsymbol{\pi}_{k}$, $T=t$, and $X=x_l$.
\end{theorem}

Using the similar arguments in Section \ref{sec:multinomial}, we define the non-negative finite sample penalty:
\eqs{
D(\boldsymbol{\delta}^{CMES}, \theta)\coloneq \sum_{k=1}^{K^L} \exp \lt( -2\Delta_{M^{*}k}^2\cdot\lt\{\sum_{l=1}^L \Pr(X=x_l)^2 (A_{kl}+A_{M^{*}l})\rt\}^{-1} \rt)\cdot \Delta_{M^{*}k},
}
and derive the following inequality:
  \begin{align}
      D(\boldsymbol{\delta}^{CMES}, \theta) \leq \frac{1}{2}\cdot e^{-\frac{1}{2}} \sum_{k=1, k\neq M^{*}}^{K^L} \lt\{\sum_{l=1}^L \Pr(X=x_l)^2 (A_{kl}+A_{M^{*}l})\rt\}^{\frac{1}{2}}.
  \end{align}
Then, we can derive the uniform bound of the regret function, which can be recovered from the observable:
\begin{align} \label{bound on maximum regret 1}
    0\leq \sup_{\theta \in \Theta}R(\boldsymbol{\delta}^{CMES}, \theta) \leq  \frac{1}{2}\cdot e^{-\frac{1}{2}} \sum_{k=1, A_{kl}\neq \bar{A}_l}^{K} \lt\{\sum_{l=1}^L \Pr(X=x_l)^2 (A_{kl}+\bar{A}_l)\rt\}^{\frac{1}{2}},
\end{align}
where $\bar{A}_l\coloneq \max_{k \in \{1,\dots, K^L \}}A_{kl}$  $ \forall l \in \mathcal{L}.$

We next investigate the relationship between the sample size and the proper conditioning level of covariates.
Recall that given a fixed sample size using all available covariates may reduce the statistical precision in practice. Let $\mathcal{Z} \coloneq \{z_1,\dots z_{L'}\}$ be a partitioning of the covariate space that is coarser than $\mathcal{X}$. Thus $L'< L$ and there exists a mapping $z(\cdot) :\mathcal{X} \mapsto \mathcal{Z}.$
Slightly abusing notation, we use the same $\boldsymbol{\pi}$ and $\boldsymbol{p}$ for assignment ratios and proportions whose dimension is $L'$. Finally, if $\boldsymbol{\pi}_{k'}$ is a generic element of   $\Pi^{L'}$ and $\boldsymbol{\delta}_Z^{CMES}$ is the MES rule conditional on $Z$, then the population expected outcome becomes:
\begin{align*}
u(\boldsymbol{\delta}_Z^{CMES}, \theta)
    &= \sum_{k'=1}^{K^{L'}} \Pr\lt(\bigcap^{K^{L'}}_{j=1, j\neq k'}\{\hat{U}(\boldsymbol{\pi}_{k'})> \hat{U}(\boldsymbol{\pi}_j) \}\rt)\cdot U(\boldsymbol{\pi}_{k'}, \theta),
\end{align*}
where
 $ U(\boldsymbol{\pi}_{k'}, \theta)
  \coloneq \sum_{l'=1}^{L'} \Pr(Z=z_{_{l'}})\cdot U_{l'}(\boldsymbol{\pi}_{k'},\theta) \equiv \sum_{l'=1}^{L'}  \Pr(Z=z_{_{l'}})\cdot[(1 - \pi_{k'l'}) \cdot E_{P_{\tht}}[Y_0(\bar{\pi}_{k'})|Z=z_{_{l'}}] + \pi_{k'l'} \cdot E_{P_{\tht}}[Y_1(\bar{\pi}_{k'})|Z=z_{_{l'}}] ]$ and  $\bar{\pi}_{k'}:= \boldsymbol{p}'\boldsymbol{\pi}_{k'} ,$
  $ k' \in \{1, \dots, K^{L'}\} $ and $ l' \in  \{1, \dots, L'\}.$

Similar to the results in Theorem \ref{thm:bounds-cmes}, we can bound the expected outcome in the following corollary.

\begin{cor}\label{cor:bounds-zcmes}
 Fix $\theta\in\Theta$. Let $\Pi^{L'} = \{\boldsymbol{\pi}_1, \dots, \boldsymbol{\pi}_{K^{L'}} \}$, $\Delta_{k'j} \coloneq|U(\boldsymbol{\pi}_{k'}, \theta)- U(\boldsymbol{\pi}_{j}, \theta)|$ for $k',j\in 1, \dots, K^{L'}$ and $\boldsymbol{\pi}_{M^{**}}:=\argmax_{\boldsymbol{\pi} \in \Pi^{L'}}  U(\boldsymbol{\pi},\theta)$.
 Then, the following inequality holds:
 \begin{align}
    &\sum_{l'=1}^{L'}  \Pr(Z=z_{_{l'}})\cdot  U_{l'}(\boldsymbol{\pi}_{M^{**}}, \theta)-  \sum_{k'=1}^{K^{L'}} \exp \lt( -2\Delta_{M^{**}k'}^2\cdot\lt\{\sum_{l'=1}^{L'} \Pr(Z=z_{_{l'}})^2 (A_{k'l'}+A_{M^{**}l'})\rt\}^{-1} \rt)\cdot \Delta_{M^{**}k'}  \nonumber\\
& \hskip140pt\leq u(\boldsymbol{\delta}_Z^{CMES}, \theta) \leq \sum_{l'=1}^{L'}  \Pr(Z=z_{_{l'}})\cdot U_{l'}(\boldsymbol{\pi}_{M^{**}}, \theta)
\end{align}
where  $A_{k'l'}\coloneq (1-\pi_{k'l'})^{2}N_{k'0l'}^{-1} + \pi_{k'l'}^{2}N_{k'1l'}^{-1} $ and $A_{{M^{**}}l'}\coloneq(1-\boldsymbol{\pi}_{l1})^{2}N_{{M^{**}}0l'}^{-1} +  \pi_{l1}^{2}N_{{M^{**}}1l'}^{-1}$, with $N_{k'tl'}$ representing the number of individuals with $\boldsymbol{\pi}_{k'}$, $Z=z_{l'}$,  and $T=t$.
\end{cor}
We now suppose that the decision maker need to choose the conditioning level between $X$ and $Z$. The idealized bounds for the regret function is as follows.
\begin{align}
    \sum_{l=1}^L  & \Pr(X=x_l)\cdot U_l(\boldsymbol{\pi}_{M^*}, \theta)-\sum_{l'=1}^{L'}  \Pr(Z=z_{_{l'}})\cdot  U_{l'}(\boldsymbol{\pi}_{M^{**}}, \theta) \nonumber
    \leq R(\boldsymbol{\delta}_Z^{CMES}, \theta)\nonumber\\
    &\leq  \sum_{l'=1}^L  \Pr(X=x_l)\cdot  U_l(\boldsymbol{\pi}_{M^*}, \theta)-\sum_{l'=1}^{L'}  \Pr(Z=z_{_{l'}})\cdot  U_{l'}(\boldsymbol{\pi}_{M^{**}}, \theta)+D(\theta) .
\end{align}
where 
$$D(\theta)\coloneq\sum_{k'=1}^{K'} \exp \lt( -2\Delta_{M^{**}k'}^2\cdot\lt\{\sum_{l'=1}^{L'} \Pr(Z=z_{_{l'}})^2 (A_{k'l'}+A_{M^{**}l'})\rt\}^{-1} \rt)\cdot \Delta_{M^{**}k'}.$$
Finally, we achieve a uniform bounds on the maximum regret function as follow.
\begin{align}\label{bound on maximum regret for zcmes}
     L \leq\sup_{\theta \in \Theta} R(\boldsymbol{\delta}_Z^{CMES}, \theta)\leq H,
\end{align}
where
$$L\coloneq\sup_{\theta \in \Theta} \left\{\sum_{l=1}^L  \Pr(X=x_l)\cdot U_l(\boldsymbol{\pi}_{M^{*}}, \theta)-\sum_{l'=1}^{L'}  \Pr(Z=z_{_{l'}})\cdot  U_{l'}(\boldsymbol{\pi}_{M^{**}}, \theta)\right\},$$
and
$$H\coloneq\sup_{\theta \in \Theta} \left\{\sum_{l=1}^L  \Pr(X=x_l)\cdot  U_l(\boldsymbol{\pi}_{M^{*}}, \theta)-\sum_{l'=1}^{L'}  \Pr(Z=z_{_{l'}})\cdot  U_{l'}(\boldsymbol{\pi}_{M^{**}}, \theta) + D(\theta)\right\}.$$
Using these bounds, we can compute the minimum sample size to test the proper level of conditioning variables. Let $\boldsymbol{N}_{KTL}:= \left(N_{ktl}: k=1,\ldots,K, t=0,1,\mbox{ and }  l=1,\ldots,L\right)$ be a 3-dimensional array of stratum sample sizes.
Recall that the upper bound of the maximum regret conditional on $X$ decreases as each $N_{ktl}$ increases.
Therefore, we can find a sufficient sample size that justifies conditioning on $X$ rather than conditioning on $Z$:
 \begin{align*}
     &\min \boldsymbol{N}_{KTL} \\
     & \hskip20pt  \mbox{subject to } L> \frac{1}{2}\cdot e^{-\frac{1}{2}} \sum_{k=1, A_{kl}\neq \bar{A}_l}^{K} \lt\{\sum_{l=1}^L \Pr(X=x_l)^2 (A_{kl}+\bar{A}_l)\rt\}^{\frac{1}{2}},
 \end{align*}
where we minimize each component of vector $\bold{N}_{TKL}$. Similar to the results in \citet{manski2004statistical}, it requires additional bound conditions on $U_l(\pi_{M^*},\theta)$ and $U_{l'}(\pi_{M^{**}},\theta)$ to solve for $\boldsymbol{N}_{KTL}$. Note also that the solution may not be unique since $\boldsymbol{N}_{KTL}$ is a tensor. 


\subsection{Numerical Experiments}

In this subsection, we conduct some numerical experiments, where we determine a sufficient sample size to use covariate-dependent treatment rules.
Suppose that we have a binary covariate $X=\{low,high\}$ available in a sample.
We now construct a treatment rule with or without the covariate.
The sufficient sample size guarantees that the maximum regret from a covariate-dependent rule is smaller than that from a rule without considering any covariate.
Thus, we can focus on covariate-dependent rules if the sample size is bigger than the sufficient one.

In this experiment, a sample is partitioned into 2 groups ($X=low$, $X=high$), and $L = \vert \mathcal{X} \vert= 2$. Therefore, covariate-dependent rules $\boldsymbol{\pi}$ also becomes a 2-dimensional vector $\boldsymbol{\pi}=(\pi_{low},\pi_{high})$.
Suppose that we consider two possible treatment rules, $\{\boldsymbol{\pi}_1=  (0.5, 0.5), \boldsymbol{\pi}_2=  (0.7, 0.3) \}.$
Unconditional treatment ratios for them becomes:
\eqs{
\bar{\pi}_1 & = \Pr(X=low)\cdot0.50 + \Pr(X=high)\cdot0.50,\\
\bar{\pi}_2 & = \Pr(X=low)\cdot0.70 + \Pr(X=high)\cdot0.30.
}
We set that $\Pr(X=low)$ varies in $\{0.1, 0.5, 0.9, 0.99\}$ and that $\Pr(X=high):=1-\Pr(X=low)$ varies in $\{0.9, 0.5, 0.1, 0.01\}$. Recall that $(N_{ktx}: k=1,2, t=0,1, \mbox{ and } x= low, high)$ denotes the sample size of each partition separated by treatment rule $k$, treatment status $t$, and covariate $x$. In addition, $N$ denote the total sample size. $N_1$ and $N_2$ denote the sample sizes of each cluster, where we apply $\boldsymbol{\pi_1}$ and $\boldsymbol{\pi_2}$, respectively.
Assuming that all states of the nature are feasible, we compute the lower bound of maximum regret for the MES rule that does not depend on covariate $X$.
We also compute the upper bounds of maximum regret for the covariate-dependent MES rule as the sample size increases. We then check when this upper bound with covariates becomes smaller than the lower bound without covariates.

In Tables \ref{table0.1}--\ref{table0.99}, we summarize the experiment results. We denote the upper bound with $X$ in bold when it becomes smaller than the lower bound without $X$.
The sufficient sample size is as low as $N=21$ when $\Pr(X=low)= 0.1$, $N=18$ when $\Pr(X=low)= 0.5,$ $N=68$ when $\Pr(X=low)= 0.9,$  and $N=5875$ when $\Pr(X=low)= 0.99.$ In each table, We also provide a breakdown of the sample sizes in each partition. This numerical study shows that covariate-dependent treatment rules can be justified with relatively small sample sizes unless the sizes of heterogeneous groups are quite uneven, e.g.\ $\Pr(X=low)=0.99$.

\begin{table}[H]
\caption{Sufficient Sample Sizes: $\Pr(X=low)=0.10$} \label{table0.1}
\centering
\resizebox{\textwidth}{!} {\begin{tabular}{rrrrrrrrrrrcc}
\\
  \hline
 \multirow{2}{*}{$N$} & \multirow{2}{*}{$N_1$} & \multirow{2}{*}{$N_2$} & \multirow{2}{*}{$N_{10low}$} & \multirow{2}{*}{$N_{11low}$} & \multirow{2}{*}{$N_{20low}$} & \multirow{2}{*}{$N_{21low}$} & \multirow{2}{*}{$N_{10high}$} & \multirow{2}{*}{$N_{11high}$} & \multirow{2}{*}{$N_{20high}$} & \multirow{2}{*}{$N_{21high}$} & Upper bound & Lower bound \\
  &  &  &  &  &  & &  &  &  &  & with $X$ & without $X$ \\
  \hline
 21 & 10 & 11 & 1 & 1 & 1 & 1 & 4 & 4 & 6 & 3      & \textbf{0.144} & 0.450 \\
 37 & 18 & 19 & 1 & 1 & 1 & 2 & 8 & 8 & 11 & 5     & \textbf{0.100} & 0.450 \\
 52 & 26 & 26 & 2 & 2 & 1 & 2 & 11 & 11 & 16 & 7   & \textbf{0.085} & 0.450 \\
 68 & 34 & 34 & 2 & 2 & 1 & 3 & 15 & 15 & 21 & 9   & \textbf{0.074} & 0.450 \\
 82 & 40 & 42 & 2 & 2 & 2 & 3 & 18 & 18 & 26 & 11  & \textbf{0.067} & 0.450 \\
 100 & 50 & 50 & 3 & 3 & 2 & 4 & 22 & 22 & 31 & 13 & \textbf{0.061} & 0.450 \\
 116 & 58 & 58 & 3 & 3 & 2 & 4 & 26 & 26 & 36 & 16 & \textbf{0.056} & 0.450 \\
 132 & 66 & 66 & 4 & 4 & 2 & 5 & 29 & 29 & 41 & 18 & \textbf{0.053} & 0.450 \\
 149 & 74 & 75 & 4 & 4 & 3 & 6 & 33 & 33 & 46 & 20 & \textbf{0.050} & 0.450 \\
 162 & 80 & 82 & 4 & 4 & 3 & 6 & 36 & 36 & 51 & 22 & \textbf{0.048} & 0.450 \\
   \hline
\end{tabular}}
\end{table}

\begin{table}[H]
\centering
\caption{Sufficient Sample Sizes: $\Pr(X=low)=0.50$}
\resizebox{\textwidth}{!} {\begin{tabular}{rrrrrrrrrrrcc}
\\
  \hline
 \multirow{2}{*}{$N$} & \multirow{2}{*}{$N_1$} & \multirow{2}{*}{$N_2$} & \multirow{2}{*}{$N_{10low}$} & \multirow{2}{*}{$N_{11low}$} & \multirow{2}{*}{$N_{20low}$} & \multirow{2}{*}{$N_{21low}$} & \multirow{2}{*}{$N_{10high}$} & \multirow{2}{*}{$N_{11high}$} & \multirow{2}{*}{$N_{20high}$} & \multirow{2}{*}{$N_{21high}$} & Upper bound & Lower bound \\
  &  &  &  &  &  & &  &  &  &  & with $X$ & without $X$ \\
  \hline
18 & 8 & 10 & 2 & 2 & 2 & 3 & 2 & 2 & 3 & 2             & \textbf{0.145} & 0.250 \\
 34 & 16 & 18 & 4 & 4 & 3 & 6 & 4 & 4 & 6 & 3           & \textbf{0.104} & 0.250 \\
 50 & 24 & 26 & 6 & 6 & 4 & 9 & 6 & 6 & 9 & 4           & \textbf{0.086} & 0.250 \\
 66 & 32 & 34 & 8 & 8 & 5 & 12 & 8 & 8 & 12 & 5         & \textbf{0.075} & 0.250 \\
 81 & 40 & 41 & 10 & 10 & 7 & 14 & 10 & 10 & 14 & 6     & \textbf{0.067} & 0.250 \\
 98 & 48 & 50 & 12 & 12 & 8 & 17 & 12 & 12 & 17 & 8     & \textbf{0.061} & 0.250 \\
 114 & 56 & 58 & 14 & 14 & 9 & 20 & 14 & 14 & 20 & 9    & \textbf{0.057} & 0.250\\
 130 & 64 & 66 & 16 & 16 & 10 & 23 & 16 & 16 & 23 & 10  & \textbf{0.053} & 0.250 \\
 146 & 72 & 74 & 18 & 18 & 11 & 26 & 18 & 18 & 26 & 11  & \textbf{0.050} & 0.250 \\
 161 & 80 & 81 & 20 & 20 & 13 & 28 & 20 & 20 & 28 & 12  & \textbf{0.048} & 0.250 \\
   \hline
\end{tabular}}
\label{table0.5}
\end{table}

\begin{table}[H]
\centering
\caption{Sufficient Sample Sizes: $\Pr(X=low)=0.90$}
\resizebox{\textwidth}{!} {\begin{tabular}{rrrrrrrrrrrcc}
\\
  \hline
 \multirow{2}{*}{$N$} & \multirow{2}{*}{$N_1$} & \multirow{2}{*}{$N_2$} & \multirow{2}{*}{$N_{10low}$} & \multirow{2}{*}{$N_{11low}$} & \multirow{2}{*}{$N_{20low}$} & \multirow{2}{*}{$N_{21low}$} & \multirow{2}{*}{$N_{10high}$} & \multirow{2}{*}{$N_{11high}$} & \multirow{2}{*}{$N_{20high}$} & \multirow{2}{*}{$N_{21high}$} & Upper bound & Lower bound \\
  &  &  &  &  &  & &  &  &  &  & with $X$ & without $X$ \\
  \hline
 21 & 10 & 11 & 4 & 4 & 3 & 6 & 1 & 1 & 1 & 1       & 0.136 & 0.072 \\
 37 & 18 & 19 & 8 & 8 & 5 & 11 & 1 & 1 & 2 & 1      & 0.100 & 0.072 \\
 52 & 26 & 26 & 11 & 11 & 7 & 16 & 2 & 2 & 2 & 1    & 0.085 & 0.072 \\
 68 & 34 & 34 & 15 & 15 & 9 & 21 & 2 & 2 & 3 & 1    & 0.074 & 0.072 \\
 82 & 40 & 42 & 18 & 18 & 11 & 26 & 2 & 2 & 3 & 2   & \textbf{0.067} & 0.072 \\
 100 & 50 & 50 & 22 & 22 & 13 & 31 & 3 & 3 & 4 & 2  & \textbf{0.061} & 0.072 \\
 116 & 58 & 58 & 26 & 26 & 16 & 36 & 3 & 3 & 4 & 2  & \textbf{0.056} & 0.072 \\
 132 & 66 & 66 & 29 & 29 & 18 & 41 & 4 & 4 & 5 & 2  & \textbf{0.053} & 0.072 \\
 149 & 74 & 75 & 33 & 33 & 20 & 46 & 4 & 4 & 6 & 3  & \textbf{0.050} & 0.072 \\
  162 & 80 & 82 & 36 & 36 & 22 & 51 & 4 & 4 & 6 & 3 & \textbf{0.048} & 0.072 \\
   \hline
\end{tabular}}
\label{table0.9}
\end{table}

\begin{table}[H]
\centering
\caption{Sufficient Sample Sizes: $\Pr(X=low)=0.99$}
\resizebox{\textwidth}{!} {\begin{tabular}{rrrrrrrrrrrcc}
\\
  \hline
 \multirow{2}{*}{$N$} & \multirow{2}{*}{$N_1$} & \multirow{2}{*}{$N_2$} & \multirow{2}{*}{$N_{10low}$} & \multirow{2}{*}{$N_{11low}$} & \multirow{2}{*}{$N_{20low}$} & \multirow{2}{*}{$N_{21low}$} & \multirow{2}{*}{$N_{10high}$} & \multirow{2}{*}{$N_{11high}$} & \multirow{2}{*}{$N_{20high}$} & \multirow{2}{*}{$N_{21high}$} & Upper bound & Lower bound \\
  &  &  &  &  &  & &  &  &  &  & with $X$ & without $X$ \\
  \hline
 21 & 10 & 11 & 4 & 4 & 3 & 6 & 1 & 1 & 1 & 1                     & 0.14609 & 0.00792 \\
 37 & 18 & 19 & 8 & 8 & 5 & 12 & 1 & 1 & 1 & 1                    & 0.10463 & 0.00792 \\
 53 & 26 & 27 & 12 & 12 & 8 & 17 & 1 & 1 & 1 & 1                  & 0.08590 & 0.00792 \\
 69 & 34 & 35 & 16 & 16 & 10 & 23 & 1 & 1 & 1 & 1                 & 0.07455 & 0.00792 \\
 84 & 42 & 42 & 20 & 20 & 12 & 28 & 1 & 1 & 1 & 1                 & 0.06721 & 0.00792 \\
 5764 & 2882 & 2882 & 1426 & 1426 & 856 & 1996 & 15 & 15 & 21 & 9 & 0.00799 & 0.00792 \\
 5780 & 2890 & 2890 & 1430 & 1430 & 858 & 2002 & 15 & 15 & 21 & 9 & 0.00798 & 0.00792 \\
 5796 & 2898 & 2898 & 1434 & 1434 & 861 & 2007 & 15 & 15 & 21 & 9 & 0.00797 & 0.00792 \\
 5812 & 2906 & 2906 & 1438 & 1438 & 863 & 2013 & 15 & 15 & 21 & 9 & 0.00796 & 0.00792 \\
 5828 & 2914 & 2914 & 1442 & 1442 & 865 & 2019 & 15 & 15 & 21 & 9 & 0.00795 & 0.00792 \\
 5844 & 2922 & 2922 & 1446 & 1446 & 868 & 2024 & 15 & 15 & 21 & 9 & 0.00793 & 0.00792 \\
 5860 & 2930 & 2930 & 1450 & 1450 & 870 & 2030 & 15 & 15 & 21 & 9 & 0.00792 & 0.00792 \\
 5875 & 2938 & 2937 & 1454 & 1454 & 872 & 2035 & 15 & 15 & 21 & 9 & \textbf{0.00791} & 0.00792 \\
 5892 & 2946 & 2946 & 1458 & 1458 & 875 & 2041 & 15 & 15 & 21 & 9 & \textbf{0.00790} & 0.00792 \\
 5907 & 2954 & 2953 & 1462 & 1462 & 877 & 2046 & 15 & 15 & 21 & 9 & \textbf{0.00789} & 0.00792 \\
5924 & 2962 & 2962 & 1466 & 1466 & 880 & 2052 & 15 & 15 & 21 & 9  & \textbf{0.00788} & 0.00792 \\
   \hline
\end{tabular}}
\label{table0.99}
\end{table}

\section{Asymptotic Optimality}\label{sec:optimality}

In this section, we study the asymptotic optimality of the multinomial empirical success (MES) rule.
We first transform the multivariate decision problem into a vector-valued binary decision problem.
Then, we show the asymptotic optimality of MES by extending the limit experiment framework in \citet{hirano2009asymptotics} into the vector-valued binary decision problem.

We first define $K(K-1)/2$-dimensional vector
\eqs{
  \delta^{VMES}_n := \left( \delta_{n,(1,2)},\ldots, \delta_{n,(k,k')},\ldots, \delta_{n,(K-1,K)} \right)',
}where $\delta_{n,(k,k')}= \mathbbm{1}(\hat{U}(\pi_k)>\hat{U}(\pi_{k'}))$.
We will call $\delta^{VMES}_n$ the vectorized multinomial empirical success (VMES) rule.\footnote{We use the subscript $n$ hereafter to distinguish a finite sample decision rule from the corresponding asymptotic one.}
Note that the VMES rule has $2^{K (K-1)/2}$ different actions while the MES rule has only $K$ actions. 
However, a set of actions by the VMES rule is always uniquely mapped into an action by the MES rule since it gives us the preference order among all $K$ actions. 
To the best of our knowledge, this is the first paper to investigate the asymptotic optimality of a multiple statistical decision problem by transforming it into a vector-valued binary decision problem.\footnote{A similar idea has been in the multiple \emph{hypothesis testing} literature for a long time, where they convert a $K$-multiple hypothesis problem into a $2^K$-finite action problem (see, e.g.\ \citet{lehmann1952testing,lehmann1957theory} and \citet{cohen2005decision}). }

We now have $J:=K (K-1) /2$ binary decision problems. Following \citet{van1991asymptotic} and \citet{hirano2009asymptotics}, we investigate the asymptotic optimality around the local alternatives.
We first restrict our attention the parametric class of $Q$ whose extension to the semiparametric class follows immediately.
Let $\mathcal{E}_n:=\{ Q^n_{{\theta}}:{\theta} \in \Theta \subset \mathbb{R}^d \}$ be a sequence of experiments, where $\Theta$ be an open subset of $\mathbb{R}^d$. We define a vector of welfare contrasts
\eqs{
  g({\theta}) := \left(g_1({\theta}), \ldots, g_J({\theta}) \right)',
}where $g_j(\theta):=U(\pi_k,{\theta}) - U(\pi_{k'},{\theta})$ is the welfare contrast between $\pi_k$ and $\pi_{k'}$.
For notational simplicity, we use $j$ for generic combination $(k,k')$, where $j=1,\ldots, J$ and $J=K(K-1)/2$.
We consider local alternatives around $\theta_0$, where $g({\theta}_0) = 0$.
This local problem is the most difficult case in the parameter space. If $g_j({\theta}_0)\neq 0$ for a given $\theta_0$, one action is strictly dominated by the other around $\theta_0$ and the decision between $(k,k')$ becomes trivial asymptotically.

We next define a loss function. We consider a loss function that is additively separable for each binary decision problem $j$:
\eq{
  L(\delta, \theta) = \sum_{j=1}^J L_j(\delta_j, \theta), \label{eq:additive loss}
}where $L_j$ is a loss function for a binary decision rule $\delta_j$ between $\pi_k$ and $\pi_{k'}$.
Specifically, we use the regret loss in this analysis:
\eqs{
  L_j(\delta_j,\theta) &:= g_j(\theta)\left[ \mathbbm{1}(g_j(\theta)>0) - \delta_j \right].
}Using the loss function in \eqref{eq:additive loss} and experiment $Q_{\theta}^n$, we define a risk function as usual:
\eq{
  R(\delta,\theta)
  & := \int_{\Omega} L(\delta(\omega^n),\theta) dQ_{\theta}^n \label{eq:risk_function} \\
  & = \sum_{j=1}^J    \int_{\Omega} L_j(\delta_j(\omega^n),\theta) dQ_{\theta}^n  \nonumber \\
  & \equiv \sum_{j=1}^J    R_j(\delta_j,\theta). \nonumber
}Note that the risk function is also additively separable.

To achieve a tractable asymptotic experiment, we assume that $Q_{\theta}$ is differentiable in quadratic mean (DQM) at $\theta_0$. For the formal definition, let $q_{\theta}$ be the density of $Q_{\theta}$ with respect to Lebesque measure $\mu$. Then, there exists a measurable function $s(\omega)$ such that, as  $h \to 0$,
\eqs{
  \int \left[ \sqrt{q_{\theta_0+h}(\omega)} - \sqrt{q_{\theta_0}(\omega)} - \frac{1}{2} h's(\omega) \sqrt{q_{\theta_0}(\omega)}  \right]^2 d\mu(\omega) = o(\Vert h \Vert^2).
}
We can usually compute $s(\omega)$ by $s = \frac{\partial \log q_{\theta}}{\partial \theta}\vert_{\theta=\theta_0}$, and the Fisher information matrix is defined as $I_0=E_{\theta_0}[ss']$.
Applying the standard local asymptotic normality arguments, we can show that the limit experiment becomes $N(\Delta|h,I_0^{-1})$, i.e.~the multivariate normal distribution with mean $h$ and variance $I_{0}^{-1}$ (see Proposition 3.1 in \citet{hirano2009asymptotics}).

We next define the corresponding loss and risk functions in the limit experiment. Recall that $g(\theta)$ is a $J\times 1$ vector of welfare contrasts with $g(\theta_0)=0$. Let $\triangledown_\theta g$ be a $J\times d$ matrix of partial derivatives of $g$ at $\theta_0$. Then, under some smoothness assumption on $g$, we have $\sqrt{n}g(\theta_0+h/\sqrt{n}) \to (\triangledown_{\theta} g)h$. Then, we observe that
\eqs{
  \sqrt{n}L_j\lt(\dt_j, \tht_0 + \frac{h}{\sqrt{n}}\rt)
  &\to (\triangledown_{\theta} g_j)h\lt[  \mathbbm{1}\lt( (\triangledown_{\theta} g_j)h >0 \rt) - \dt_j \rt] \\
  & \equiv L_{j,\infty} (\dt_j, h),
}
where $\triangledown_{\theta} g_j$ is the $j$-th row of matrix $\triangledown_{\theta} g$. Using the additive separability, we can define the asymptotic loss function as
\eqs{
  L_{\infty}(\dt,h) := \sum_{j=1}^J L_{j,\infty} (\dt_j, h).
}Similarly, we can define the corresponding asymptotic risk function as
\eqs{
  R_{j,\infty}(\dt_j,h) & := \lim_{n\to\infty} \sqrt{n} R_j\lt(\dt_j,\tht_0+\frac{h}{\sqrt{n}}\rt)\\
  & = \int L_{j,\infty}(\dt_j(\Delta),h)dN(\Delta|h,I_0^{-1})\\
  R_{\infty}(\dt, h) & := \sum_{j=1}^J R_{j,\infty}(\dt_j,h)\\
  R_{\infty}(\dt) & := \sup_{h \in \mathbb{R}^d} R_{\infty}(\dt,h).
}Abusing notation slightly, we use the same $\dt_j$ for both $R_{j}(\dt_j,\theta)$ and  $R_{j,\infty}(\dt_j,h)$. However, notice that one in $R_j$ is defined on the sample sample $\omega^n \in \Omega$ while the other in $L_{j,\infty}$ is on the simpler asymptotic experiment space $\Delta \in \mathbb{R}^d$.

In the next theorem we characterize the functional minimization problem in the limiting Gaussian experiment. 
We first define additional notation. Let $h_0$ be a vector such that $(\triangledown_{\theta} g) h_0 = 0$. For any $b_j \in \mathbb{R}$, we slice the parameter space as follows
\eqs{
  h_j(b_j,h_0) = h_0 + \frac{b_j}{(\triangledown_{\theta} g_j) I_0^{-1} (\triangledown_{\theta} g_j)'} I_0^{-1} (\triangledown_{\theta} g_j)'.
}Note that parameter $h_j(b_j,h_0)$ in the slice satisfies $(\triangledown_{\theta} g_j)h_j = b_j$, which is the $j$-th component of the welfare contrast vector $g(\theta)$. Because of the additive risk function and the Neyman-Pearson lemma, we can characterize the minimization problem by investigating a set of threshold rules over a vector of the sliced parameter space, separately.

\begin{theorem}\label{thm:slicing}
Let $\Delta\sim N(h,I_0^{-1})$ for $h\in \mathbb{R}^d$ and $L_{\infty}(h):=diag(L_{j,\infty}(1,h)-L_{j,\infty}(0,h))$ be the $(J\times J)$ diagonal matrix whose $(j,j)$ element is $L_{j,\infty}(1,h)-L_{j,\infty}(0,h)$. Consider a simple finite action problem $(a_1,\ldots,a_J)$ with $a_j \in\{0,1\}$. For all $h$ with $(\triangledown_{\theta} g)h \neq 0$, loss functions $\{L_{j,\infty}(a,h)\}$ satisfies that
\eq{
  L_{\infty}(h) ((\triangledown_{\theta} g) h) < 0, \label{eq:right_loss}
}where the inequality holds element-by-element.

\begin{enumerate}
  \item[(i)] Let $\tilde{\boldsymbol{\delta}}(\Delta)$ be a $(J \times 1)$ vector of any randomized decision rules. Let $h_0 \in\mathbb{R}^d$ be given. Suppose that risk function $R_{\infty}(\delta,h)$ is additively separable, i.e.\ $R_{\infty}(\delta,h)=\sum_{j=1}^J R_{j,\infty}(\delta_j,h)$. Then, there exists a rule $ \boldsymbol{\delta}_c :=\left(\delta_{1,c_1}(\Delta), \ldots,\delta_{J,c_J}(\Delta)  \right)'$  with $\delta_{j,c_j}= \mathbbm{1}\left(  (\triangledown_{\theta} g_j)  \Delta > c_j)\right)$ for $j=1,\ldots,J$ such that
  \eqs{
  R_{\infty}(\boldsymbol{\delta}_c(\Delta), h)  \le R_{\infty}(\tilde{\boldsymbol{\delta}}(\Delta), h),
  }on the subspace $\{h_j(b_j,h_0):b_j \in \mathbb{R}, j=1,\ldots,J\}$.
  \item[(ii)] Suppose that $L_{j,\infty}(a_j,h)$ depends on $h$ only through $ (\triangledown_{\theta} g_j) h$ for all $j$. If there exists a minimax rule, then $\boldsymbol{\delta}_{c^*}(\Delta)$ is minimax for some $(J\times1)$ vector $c^*$. The optimal vector $c_j^*$ can be achieved by solving $\inf_{c_j} \sup_{b} E_{h_j(b_j,0)}L_{\infty}( \boldsymbol{\delta}_{c_j} , h_j(b_j,0))$ for all $j=1,\ldots,J$.
\end{enumerate}
\end{theorem}
Condition \eqref{eq:right_loss} requires that higher loss be assigned to any incorrect choice for each $j$, and $L_{j,\infty}(\delta_j,h)$ clearly satisfies the condition. Since the expectation is a linear operator the additively separable loss function in \eqref{eq:additive loss} assures the additive separability of risk function $R_{\infty}(\delta,h)$. Theorem \ref{thm:slicing} (i) implies that threshold rule $\delta_{c}(\Delta)$ is \emph{admissible} on the subspace. This result is an extension of the Theorem 3.4 in \citet{hirano2009asymptotics} into a finite action problem.

To finalize our arguments on the minimax optimality, we collect all the regularity conditions.
\begin{assum}\label{ass:function_g}
Let $g$ be a vector-valued welfare contrast function whose dimension is $J\times 1$. Then, it satisfies that $g(\theta_0)=0$ and $g(\theta)$ is differentiable at $\theta_0$.
\end{assum}

\begin{assum}\label{ass:LAN}
The sequence of experiments $\mathcal{E}_n:=\{Q^n_{\tht}: \theta\in\Theta, n=1,\ldots\}$ is differentiable in quadratic mean at $\theta_0 \in \Theta \subset \mathbb{R}^d$ with nonsingular $I_0$.
\end{assum}

\begin{assum}\label{ass:estimator}
(i) There exists the best regular estimator $\hat{\tht}$ such that
\eqs{
  \sqrt{n}(\hat{\tht}_n-\tht_0 -h/\sqrt{n}) \overset{h}{\rightsquigarrow} N(0,I_0^{-1})~~ \forall h \in \mathbb{R}^d,
}where $\overset{h}{\rightsquigarrow}$ denotes the convergence in distribution under the sequence of $Q_{\theta_0 + h/\sqrt{n}}^n$.\\
(ii) Let $\sigma_{g_j}^2:=(\triangledown_\theta g_j) I_0^{-1} (\triangledown_\theta g_j)'$. Then, there exists an estimator $\hat{\sigma}_{g_j}$ such that
\eqs{
  \hat{\sigma}_{g_j} \overset{p}{\to} \sigma_{g_j}~~\forall j =1,\ldots,J
}under $\theta_0$.
\end{assum}

These regularity conditions are similar to those in \citet{hirano2009asymptotics}. Assumption \ref{ass:function_g} is a mild extension of the welfare contrast to a vector-valued function. We impose that the smoothness assumption holds element-by-element. Assumption \ref{ass:LAN} is the standard condition for the local asymptotic normality. Therefore, the asymptotic experiment can be approximated by the multivariate normal distribution for each $j$. Finally, Assumption \ref{ass:estimator} assures the existence of an efficient estimator for $\theta_0$ and a consistent estimator for $\sigma_{g_j}$ for each $j$.

\begin{theorem}\label{thm:asymp_opt_parametric}
Suppose that Assumptions \ref{ass:function_g}--\ref{ass:estimator} hold. Let $\dt_n^R$ be a $J \times 1$ dimensional decision rule whose $j$-th component is defined as
\eqs{
  \dt_{n,j}^R := \mathbbm{1} \lt(\frac{\sqrt{n}g_j(\hat{\tht}_n)}{\hat{\sigma}_{g_j}}>0\rt),
}
Then, it holds that
\eq{
  \sup_{H \in \mathcal{H}} \liminf_{n\to \infty}\sup_{h \in H} \sqrt{n} R\lt(\delta_n^{R}, \theta_0 + \frac{h}{\sqrt{n}} \rt) =
  \inf_{\delta_n\in\mathcal{D}} \sup_{H \in \mathcal{H}} \liminf_{n \to \infty} \sup_{h \in H} \sqrt{n} R \lt(\dt_n, \theta_0 +\frac{h}{\sqrt{n}}\rt),
}where $\mathcal{H}$ is a collection of all finite subsets of $\mathbb{R}^d$ and $\mathcal{D}$ is the set of all sequences of decision rules that converges to the asymptotic decision problem.
\end{theorem}

This theorem is a gentle extension of Theorem 3.5 of \citet{hirano2009asymptotics} to the finite action framework with the additively separable loss function. In this paper, we focus on the statistical decision problem under social interaction, where it is transformed into choosing the fraction of the treatment. However, the result of this theorem is applicable to any case, where the decision problem is represented as a choice among multiple actions.

Straightforward is an extension to semiparametric models. We have restricted our attention to the class of parametric models $\Theta$ in this section, but we can extend it to the class of distributions $\mathcal{P}$ with more complicated notation. Instead of repeating the same arguments with messier notation, we refer to \citet[section 4]{hirano2009asymptotics} and \citet{van1991asymptotic} for the extension. The main difference is that the multivariate Gaussian limit experiment is now replaced by an infinite Gaussian sequence.

Since the optimal decision rule has the same threshold constant both in parametric models and semiparametric models, we can claim the asymptotic optimality of the MES rule based on the finite action framework. Suppose that we have a random sample $(y_t(\pi_k), y_t(\pi_{k'}))$ for the binary decision problem between $\pi_k$ and $\pi_{k'}$. Let $F_{t,k}$ and $F_{t,k'}$ be the distributions of the sample, which is unknown but included in the class of $\mathcal{P}$. We assume that $\mathcal{P}$ is the largest class of distributions satisfying
\eqs{
  \sup_{F\in\mathcal{P}} \int \vert y \vert^2 dF(y) < \infty.
}Recall that the welfare contrast function in this binary decision problem becomes
\eqs{
  g_j & = U(\pi_k) - U(\pi_{k'}) \\
      & = (1-\pi_k) \int y dF_{0,k}(y) + \pi_k \int y dF_{1,k}(y)
      -(1-\pi_{k'}) \int y dF_{0,k'}(y) + \pi_{k'} \int y dF_{1,k'}(y).
}Note that the MES rule can be written as
\eqs{
  \delta_{n,(k,k')} & =  \mathbbm{1} ( \hat{g}_{n,j} > 0 ),
}where
\eqs{
  \hat{g}_{n,j}
    & := \hat{U}(\pi_k) - \hat{U}(\pi_{k'}) \\
    & = (1-\pi_k)\cdot \frac{\sum_{n_k=1}^{N_{k}} y_{n_k}(\pi_k)\cdot  \mathbbm{1}(t_{n_k}(\pi_k)=0)}{\sum_{n_k=1}^{N_{k}}   \mathbbm{1}(t_{n_k}(\pi_k)=0)}   +  \pi_k \cdot  \frac{\sum_{n_k=1}^{N_{k}} y_{n_k}(\pi_k)\cdot  \mathbbm{1}(t_{n_k}(\pi_k)=1)}{ \sum_{n_k=1}^{N_{k}}  \mathbbm{1}(t_{n_k}(\pi_k)=1)} \\
    & \hskip30pt - (1-\pi_{k'})\cdot \frac{\sum_{n_{k'}=1}^{N_{k'}} y_{n_{k'}}(\pi_{k'})\cdot  \mathbbm{1}(t_{n_k'}(\pi_{k'})=0)}{\sum_{n_{k'}=1}^{N_{k'}}   \mathbbm{1}(t_{n_{k'}}(\pi_{k'})=0)}   +  \pi_{k'} \cdot  \frac{\sum_{n_{k'}=1}^{N_{k'}} y_{n_k'}(\pi_{k'})\cdot  \mathbbm{1}(t_{n_k'}(\pi_{k'})=1)}{ \sum_{n_{k'}=1}^{N_{k'}}  \mathbbm{1}(t_{n_k'}(\pi_{k'})=1)}.
}Since $\hat{g}_{n,j}$ is an asymptotically efficient estimator of $g_j$ \citep{bickel1993efficient}, we can conclude that $\delta_{n,(k,k')}$ is asymptotically minimax optimal for the regret loss function and that the MES rule is asymptotically optimal for the additively separable loss function.

\section{Conclusion}\label{sec:conclusion}
In this paper we study statistical treatment rules under social interaction. We impose the anonymous interaction assumption, and consider a treatment decision problem, where we choose the treatment ratio for each cluster. We propose a simple but intuitive rule called the multinomial empirical success (MES) rule. We construct the finite sample regret bound of the MES rule and show how it can be applied in the treatment decision problems. Finally, we show that the proposed MES rule achieves the asymptotic optimality in the sense of \citet{hirano2009asymptotics}.

We may consider a few possible extensions. It is interesting to investigate the finite sample optimality of the MES rule. It does not work immediately if we apply the finite action problem framework, which we adopt in the asymptotic optimality analysis, and the game theoretic approach in \citet{stoye2009minimax} in the finite sample case. It is also interesting to relax the anonymous interaction assumption. Then, we have to ask what kind of additional information help reduce the dimension of the action space. The network information can be such an example. We leave these questions for our future research.

\newpage
\appendix

\section*{Appendix}

\paragraph{Proof of Theorem \ref{thm:bounds}}
We first show the bounds in the main text.
Without loss of generality, suppose  $\max_{\pi \in \mathbf{\Pi}}  U(\pi,\theta) =  U(\pi_1,\theta)$. 
By definition, the upper bound of $u(\delta^{MES}, \theta)$ is $U(\pi_1,\theta)$. 
We now restate the expected welfare under the MES rule as
\begin{small}
   \begin{align}\label{firsteq_1}
u(\delta^{MES}, \theta)  & =\Pr\lt(\hat{U}(\pi_1,\theta)>\max_{\pi \in \mathbf{\Pi}_{-1} } \hat{U}(\pi, \theta)\rt)\cdot U(\pi_1, \theta) + \sum_{k=2}^K \Pr\lt(\hat{U}(\pi_k,\theta)>\max_{\pi \in \mathbf{\Pi}_{-k} } \hat{U}(\pi, \theta)\rt)\cdot U(\pi_k, \theta) \nonumber \\
                              & = \Bigg[1- \sum_{k=2}^K \Pr\lt(\hat{U}(\pi_k,\theta)>\max_{\pi \in \mathbf{\Pi}_{-k} } \hat{U}(\pi, \theta)\rt)\Bigg]\cdot U(\pi_1, \theta) \nonumber\\
                              & \hskip20pt + \sum_{k=2}^K \Pr\lt(\hat{U}(\pi_k,\theta)>\max_{\pi \in \mathbf{\Pi}_{-k} } \hat{U}(\pi, \theta)\rt)\cdot U(\pi_k, \theta)
\end{align}
\end{small}

Define 
\begin{small}
\begin{align*}
w(\delta^{MES}, \theta)& \coloneq u(\delta^{MES}, \theta) - \Bigg\{  \left[  1-\sum_{k=2}^{K} \Pr(\hat{U}(\pi_k,\theta)-\hat{U}(\pi_1, \theta)> 0)\right]\cdot U(\pi_1, \theta)  \\
& \hskip20pt + \sum_{k=2}^{K} \left[ \Pr(\hat{U}(\pi_k,\theta)-\hat{U}(\pi_1, \theta)> 0)\right]\cdot U(\pi_k, \theta)\Bigg\}.
\end{align*}
\end{small}
Plugging in  (\ref{firsteq_1}) into $w$, we have
\allowdisplaybreaks
\begin{small}
   \begin{align*}
   w(\delta^{MEG}, \theta) & = \Bigg\{\Bigg[ 1- \sum_{k=2}^K \Pr\lt(\hat{U}(\pi_k,\theta)>\max_{\pi \in \mathbf{\Pi}_{-k} } \hat{U}(\pi, \theta)\rt)\Bigg] - \Bigg[1- \sum_{k=2}^{K}  \Pr\lt(\hat{U}(\pi_k,\theta)-\hat{U}(\pi_1, \theta) > 0\rt) \Bigg]\Bigg\} \cdot U(\pi_1, \theta) \\
   & \hskip20pt + \sum_{k=2}^K \Bigg[\Pr\lt(\hat{U}(\pi_k,\theta)>\max_{\pi \in \mathbf{\Pi}_{-k} } \hat{U}(\pi, \theta)\rt)-  \Pr\lt(\hat{U}(\pi_k,\theta)-\hat{U}(\pi_1,\theta)> 0\rt)\Bigg]\cdot U(\pi_k, \theta)\\
 & =  \sum_{k=2}^{K} \Bigg[ \Pr\lt(\hat{U}(\pi_k,\theta)-\hat{U}(\pi_1, \theta)> 0\rt)- \Pr\lt(\hat{U}(\pi_k,\theta)>\max_{\pi \in \mathbf{\Pi}_{-k} } \hat{U}(\pi, \theta)\rt)\Bigg]\cdot U(\pi_1, \theta)  \\
     & \hskip20pt + \sum_{k=2}^K \Bigg[\Pr\lt(\hat{U}(\pi_k,\theta)>\max_{\pi \in \mathbf{\Pi}_{-k} } \hat{U}(\pi, \theta)\rt)-  \Pr\lt(\hat{U}(\pi_k,\theta)-\hat{U}(\pi_1,\theta)> 0\rt)\Bigg]\cdot U(\pi_k, \theta) \\
      &=   \sum_{k=2}^{K} \Bigg[ \Pr\lt(\hat{U}(\pi_k,\theta)-\hat{U}(\pi_1, \theta)> 0\rt)- \Pr\lt(\hat{U}(\pi_k,\theta)>\max_{\pi \in \mathbf{\Pi}_{-k} } \hat{U}(\pi, \theta)\rt)\Bigg]\cdot U(\pi_1, \theta)  \\
     & \hskip20pt - \sum_{k=2}^K \Bigg[ \Pr\lt(\hat{U}(\pi_k,\theta)-\hat{U}(\pi_1,\theta)> 0\rt)-\Pr\lt(\hat{U}(\pi_k,\theta)>\max_{\pi \in \mathbf{\Pi}_{-k} } \hat{U}(\pi, \theta)\rt) \Bigg]\cdot U(\pi_k, \theta) \\
         & =  \sum_{k=2}^{K} \Bigg[ \Pr\lt(\hat{U}(\pi_k,\theta)-\hat{U}(\pi_1, \theta)> 0\rt)- \Pr\lt(\hat{U}(\pi_k,\theta)>\max_{\pi \in \mathbf{\Pi}_{-k} } \hat{U}(\pi, \theta)\rt)\Bigg]\cdot \lt(U(\pi_1, \theta) - U(\pi_k, \theta) \rt) \\
         & \geq 0.
\end{align*}
\end{small}
\normalsize Note that the last inequality holds since 
$$\Pr\lt(\hat{U}(\pi_k,\theta)-\hat{U}(\pi_1, \theta)\geq 0\rt)\geq\Pr\lt(\hat{U}(\pi_k,\theta)>\max_{\pi \in \mathbf{\Pi}_{-k} } \hat{U}(\pi, \theta)\rt),\, \forall k.$$

Therefore, we have
\begin{align}\label{eq19}
 u(\delta^{MEG}, \theta) & \geq   \Bigg[  1-\sum_{k=2}^{K} Pr(\hat{U}(\pi_k,\theta)-\hat{U}(\pi_1, \theta)\geq 0)\Bigg]\cdot U(\pi_1, \theta)  \notag \\
 &\hskip20pt + \sum_{k=2}^{K} \Bigg[ Pr(\hat{U}(\pi_k,\theta)-\hat{U}(\pi_1, \theta)\geq 0)\Bigg]\cdot U(\pi_k, \theta)
\end{align}

To proceed, we use the Hoeffding inequality to derive bounds for the probabilities in   (\ref{eq19}).
For $k=2,\ldots, K$,
\begin{align*}
\hat{U}(\pi_{k}, \theta)-\hat{U}(\pi_1,\theta) & =\frac{1}{N_{k} +N_{1}}\Bigg\{ \sum_{n\in N(0,\pi_k)} (1-\pi_k)y_n\frac{N_{k} +N_{1}}{N_{k0}} +\sum_{n\in N(1,\pi_k)} \pi_k y_n\frac{N_{k} +N_{1}}{N_{k1}} +\\ &\sum_{n\in N(0,\pi_1)} -(1-\pi_1)y_n\frac{N_{k} +N_{1}}{N_{10}} +
 \sum_{n\in N(1,\pi_1)} -\pi_1 y_n\frac{N_{k} +N_{1}}{N_{11}} \Bigg\}
\end{align*}
Thus,  $\hat{U}(\pi_{k}, \theta)-\hat{U}(\pi_1,\theta)$  is the average of $(N_{1} +N_{k})$ independent random variables whose ranges are   $ [0, (1-\pi_k)(N_{1} +N_{k})/N_{k0}]$ , $ [0, \pi_k (N_{1} +N_{k})/N_{k1}]$, $ [- (1-\pi_1)(N_{1} +N_{k})/N_{10}, 0]$,  and  $ [-\pi_1 (N_{1} +N_{k})/N_{11},0]$. Since   $\max_{\pi \in \mathbf{\Pi}} U(\pi, \theta)= U(\pi_1,\theta)$,  $\mathbbm{E}[\hat{U}(\pi_{k}, \theta)-\hat{U}(\pi_1,\theta)]= -|U(\pi_k, \theta)- U(\pi_1,\theta)|= -\Delta_{1k}$.  
Applying the Hoeffding inequality for all $k \neq 1$, we have
\begin{align}\label{eq20}
\Pr(\hat{U}(\pi_k,\theta)-\hat{U}&(\pi_1,\theta)\geq 0) \notag  \\
&=\Pr(\hat{U}(\pi_k,\theta)-\hat{U}(\pi_1,\theta) + \Delta_{k1} \geq \Delta_{k1}) \notag \\
& \leq  \exp \lt(-2\Delta_{k1}^2\cdot \Bigg\{(1-\pi_k)^{2}N_{k0}^{-1} + \pi_{k}^{2}N_{k1}^{-1} +
 (1-\pi_1)^{2}N_{10}^{-1} +  \pi_{1}^{2}N_{11}^{-1} \Bigg\}^{-1}\rt) \notag \\
& \equiv  \exp \lt( -2\Delta_{k1}^2\cdot (A_{k}+A_1)^{-1} \rt)
\end{align}

Substituting   (\ref{eq20})  into the last inequality of (\ref{eq19}), we obtain;
 \begin{align}\label{eq21}
    u(\delta^{MEG}, \theta) & \geq  \Bigg(1- \sum_{k=2}^{K}\exp \lt( -2\Delta_{k1}^2\cdot (A_{k}+A_1)^{-1} \rt)\Bigg)\cdot U(\pi_1, \theta) \notag \\
    &\hskip20pt 
    +  \sum_{k=2}^{K} \exp \lt( -2\Delta_{k1}^2\cdot (A_{k}+A_1)^{-1} \rt) \cdot U(\pi_k, \theta) \notag \\
    & =U(\pi_1, \theta)- \sum_{k=1}^{K} \exp \lt( -2\Delta_{k1}^2\cdot (A_{k}+A_1)^{-1} \rt)\cdot \Delta_{1k} 
\end{align}
as required when  $M^*=1$.



\hfill $\square$

\paragraph{Proof of Theorem \ref{thm:bounds-cmes} }
Without loss of generality, let $  \max_{\boldsymbol{\pi}\in \Pi^L}\sum_{l=1}^L P(X=x_l)U_l(\boldsymbol{\pi},\theta)=U(\boldsymbol{\pi}_{1},\theta).$
The upper bound is straight forward; the highest attainable outcome for the CMES rule which conditions on all covariates is
\begin{equation*}
   \max_{\boldsymbol{\pi}\in \Pi^L}\sum_{l=1}^L P(X=x_l)U_l(\boldsymbol{\pi},\theta)=U(\boldsymbol{\pi}_{1},\theta)
\end{equation*}

Now, restate the expected outcome under the CMES rule as;

\begin{small}
   \begin{align}\label{firsteq}
u(\boldsymbol{\delta}^{CMES}, \theta)  & =\Pr\lt(\hat{U}(\boldsymbol{\pi}_1,\theta)>\max_{\boldsymbol{\pi} \in \Pi^{L}_{-1} } \hat{U}(\boldsymbol{\pi}, \theta)\rt)\cdot U(\boldsymbol{\pi}_1, \theta) + \sum_{k=2}^K \Pr\lt(\hat{U}(\boldsymbol{\pi}_k,\theta)>\max_{\boldsymbol{\pi} \in \Pi^{L}_{-k} } \hat{U}(\boldsymbol{\pi}, \theta)\rt)\cdot U(\boldsymbol{\pi}_k, \theta) \notag \\
                               & =  \Bigg[1- \sum_{k=2}^K \Pr\lt(\hat{U}(\boldsymbol{\pi}_k,\theta)>\max_{\boldsymbol{\pi} \in \Pi^{L}_{-k} } \hat{U}(\boldsymbol{\pi}, \theta)\rt)\Bigg]\cdot U(\boldsymbol{\pi}_1, \theta) \notag \\
                               &\hskip20pt + \sum_{k=2}^K \Pr\lt(\hat{U}(\boldsymbol{\pi}_k,\theta)>\max_{\boldsymbol{\pi} \in \Pi^{L}_{-k} } \hat{U}(\boldsymbol{\pi}, \theta)\rt)\cdot U(\boldsymbol{\pi}_k, \theta) 
\end{align}
\end{small}
Using the same arguments as in the proof of theorem \ref{thm:bounds}, we can show that
\begin{align} \label{lower bound on EO}
 u(\boldsymbol{\delta}^{CMES}, \theta) & \geq   \Bigg[  1-\sum_{k=2}^{K} \Pr(\hat{U}(\boldsymbol{\pi}_k,\theta)-\hat{U}(\boldsymbol{\pi}_1, \theta) > 0)\Bigg]\cdot U(\boldsymbol{\pi}_1, \theta) \notag \\
 & \hskip20pt + \sum_{k=2}^{K} \Bigg[ \Pr(\hat{U}(\boldsymbol{\pi}_k,\theta)-\hat{U}(\boldsymbol{\pi}_1, \theta) > 0)\Bigg]\cdot U(\boldsymbol{\pi}_k, \theta)
\end{align}

Now, for $k=2,\ldots, K$,
\begin{align*}
\hat{U}(\boldsymbol{\pi}_{k}, \theta)-\hat{U}(\boldsymbol{\pi}_1,\theta) & =\sum_{l=1}^L\frac{P(X=x_l)}{N_{k} +N_{1}}\Bigg\{ \sum_{n\in N(0,\boldsymbol{\pi}_k, x_l)} (1-\pi_{lk})y_n\frac{{N_{k} +N_{1}}}{N_{k0l}} +\sum_{n\in N(1,\boldsymbol{\pi}_k, x_l)} \boldsymbol{\pi}_{lk} y_n\frac{{N_{k} +N_{1}}}{N_{k1l}} +\\ &\sum_{n\in N(0,\boldsymbol{\pi}_1, x_l)} (-(1-\boldsymbol{\pi}_{l1}) )y_n\frac{{N_{k} +N_{1}}}{N_{10l}} +
 \sum_{n\in N(1,\boldsymbol{\pi}_1, x_l)} (-\boldsymbol{\pi}_1) y_n\frac{{N_{k} +N_{1}}}{N_{11l}} \Bigg\}.
\end{align*}
Thus,  $\hat{U}(\boldsymbol{\pi}_{k}, \theta)-\hat{U}(\boldsymbol{\pi}_1,\theta)$  is the average of $N_{k} +N_{1}$ independent random variables whose ranges are   $ [0,P(X=x_l) (1-\pi_{lk})(N_{k} +N_{1})/N_{k0l}]$ , $ [0, P(X=x_l)\pi_{lk} (N_{k} +N_{1})/N_{k1l}]$, $ [- P(X=x_l)(1-\pi_{l1})(N_{k} +N_{1})/N_{10l}, 0]$,  and  $ [-P(X=x_l)\pi_{l1} (N_{k} +N_{1})/N_{11l},0]$.

For all $k \neq 1$, the Hoeffding inequality yields
\begin{align}\label{Hoefding result}
Pr&(\hat{U}(\boldsymbol{\pi}_k,\theta)-\hat{U}(\boldsymbol{\pi}_1,s)\geq 0) =Pr(\hat{U}(\boldsymbol{\pi}_k,\theta)-\hat{U}(\boldsymbol{\pi}_1,\theta) + \Delta_{k1} \geq \Delta_{k1}) \notag \\
& \leq  \exp \lt(-2\Delta_{k1}^2\cdot \Bigg\{\sum_{l=1}^L P(X=x_l)^2[(1-\pi_{lk})^{2}N_{k0l}^{-1} + \pi_{lk}^{2}N_{k1l}^{-1} +
 (1-\boldsymbol{\pi}_{l1})^{2}N_{10l}^{-1} +  \pi_{l1}^{2}N_{11l}^{-1}] \Bigg\}^{-1}\rt) \notag \\
& \equiv \exp \lt( -2\Delta_{k1}^2\cdot\lt\{\sum_{l=1}^L P(X=x_l)^2 (A_{kl}+A_{1l})\rt\}^{-1} \rt)
\end{align}
Plugging in (\ref{Hoefding result}) into (\ref{lower bound on EO}), we obtain

\begin{equation}
\begin{split}
 u(\boldsymbol{\delta}^{CMES}, \theta) &\geq   \Bigg[  1-\sum_{k=2}^{K}  \exp \lt( -2\Delta_{k1}^2\cdot\lt\{\sum_{l=1}^L P(X=x_l)^2 (A_{kl}+A_{1l})\rt\}^{-1} \rt)\Bigg]\cdot U(\boldsymbol{\pi}_1, \theta)\\  &+ \sum_{k=2}^{K} \Bigg[ \exp \lt( -2\Delta_{k1}^2\cdot\lt\{\sum_{l=1}^L P(X=x_l)^2 (A_{kl}+A_{1l})\rt\}^{-1} \rt)\Bigg]\cdot U(\boldsymbol{\pi}_k, \theta)  \\
 & =U(\boldsymbol{\pi}_1, \theta)- \sum_{k=1}^{K} \exp \lt( -2\Delta_{k1}^2\cdot\lt\{\sum_{l=1}^L P(X=x_l)^2 (A_{kl}+A_{1l})\rt\}^{-1} \rt)\cdot \Delta_{1k} \\
\end{split}
\end{equation}
as required when  $M^*=1$.
\hfill $\square$

\paragraph{Proof of Theorem \ref{thm:slicing}}
(i) Since the risk function is additively separable, we have
\eqs{
  R_{\infty}(\tilde{\delta}(\Delta), h) - R_{\infty}({\delta}_c(\Delta), h)
    & = \sum_{j=1}^J \left( R_{j,\infty}(\tilde{\delta}_j(\Delta), h) - R_{j,\infty}({\delta}_{j,c_j}(\Delta), h) \right).
}Thus, it is sufficient to show that
\eq{
R_{j,\infty}(\tilde{\delta}_j(\Delta), h) - R_{j,\infty}({\delta}_{j,c_j}(\Delta), h) \ge 0~~\mbox{for all $j$.} \label{eq:main equation of the admissibility thm}
}

Recall that $\triangledown_{\theta} g_j$ is the $j$-th row of the $(J \times d)$ matrix $\triangledown_{\theta} g$. Since $\triangledown_{\theta} g_j \Delta \sim N(0, \triangledown_{\theta} g_j I_0^{-1} \triangledown_{\theta} g_j')$, we can compute $E_{h_0}[\delta_{j,c_j}(\Delta)] = 1 - \Phi\lt(\Delta \le c_j/\sqrt{\triangledown_{\theta} g_j I_0^{-1} \triangledown_{\theta} g_j'}\rt)$. For any given $\tilde{\delta}_j(\Delta)$, we can set $c_j$ such that $E_{h_0}[\delta_{j,c_j}(\Delta)]=E_{h_0}[\tilde{\delta}_{j}(\Delta)]$.

Let $b_j>0$ be given. Consider the simple hypotheses test between $H_{0j}:h=h_0$ and $H_{1j}: h=h_j(b_j,h_0)$ based on $\Delta$. The Neyman-Pearson lemma implies that the most powerful test rejects $H_0$ for large values of
\eqs{
  \log\frac{dN(h_1,I_0^{-1})}{dN(h_0,I_0^{-1})}(\Delta)=\frac{b_j}{\triangledown_{\theta} g_jI_{0}^{-1}\triangledown_{\theta} g_j'}\triangledown_{\theta} g_j\Delta-\frac{1}{2}\frac{b_j^{2}}{\triangledown_{\theta} g_jI_{0}^{-1}\triangledown_{\theta} g_j'},
}which is equivalent to large values of $\triangledown_{\theta} g_j\Delta$. Therefore, we have $E_{h_j(b_j,h_0)}[{\delta}_{j,c_j}(\Delta)] \ge E_{h_j(b_j,h_0)}[\tilde{\delta}_j(\Delta)]$, which holds for all $b_j\ge0$. Similarly, we can show that $1-E_{h_j(b_j,h_0)}[{\delta}_{j,c_j}(\Delta)] \ge 1-E_{h_j(b_j,h_0)}[\tilde{\delta}_j(\Delta)]$ for all $b_j<0$, which is equivalent to $E_{h_j(b_j,h_0)}[{\delta}_{j,c_j}(\Delta)] \le E_{h_j(b_j,h_0)}[\tilde{\delta}_j(\Delta)]$. From $R_{j,\infty}(\tilde{\delta}_j(\Delta), h) - R_{j,\infty}({\delta}_{j,c_j}(\Delta), h)=(L_{j,\infty}(1,h)-L_{j,\infty}(0,h))(E_h[\tilde{\delta}_j(\Delta)]-E_h[{\delta}_{j,c_j}(\Delta)])$, condition \eqref{eq:right_loss}, and $\triangledown_{\theta} g_j h_j = b_j$, we conclude that \eqref{eq:main equation of the admissibility thm} holds for the subspace  $\{h_j(b_j,h_0):b_j\in \mathbb{R}\}$. Repeating this procedure for all $j$, we can establish the desired result.

(ii) Let $R^*:=\inf_{\delta \in \mathcal{D}_{\infty}}\sup_h R_{\infty}(\delta,h)$ the optimal minimax regret and $\delta^*$ be a solution so that $\sup_hR_{\infty}(\delta^* , h)=R^*$. Then, we have
\eqs{
  R^*   & \ge \sup_{h} R_{\infty}(\delta^*,h) \\
        & = \sup_{b} \sum_{j=1}^J R_{j,\infty}(\delta_{j}^*, h_j(b_j,0)) \\
        & \ge \sup_{b} \sum_{j=1}^J R_{j,\infty}(\delta_{j,c^*}, h_j(b_j,0)) \\
        & = \sup_{h} R_{\infty}(\delta_{c^*}, h) \\
        & \ge R^*.
}Note that the first inequality holds by definition of $\delta^*$. The second inequality holds from the additive separability and the partition property of $\{h_j(b_j,0)\}$, i.e.\ each $h$ is uniquely determined by intersection of slices. The third inequality holds from the result in (i). To see the validity of the fourth equality, note that
\eqs{
  \sum_{j=1}^J R_{j,\infty}(\delta_{j,c_j^*}, h_{j}(b_j,0))
                          & = \sum_{j=1}^J \left[ L_{j,\infty}(0,b_j) + E_{h_j(b_j,0)}\left[\delta_{j,c_j^*}(\Delta)  \right]  (L_{j,\infty}(1,b_j)-L_{j}(0,b_j)) \right] \\
                          & = \sum_{j=1}^J R_{j,\infty}(\delta_{j,c_j^*}, h_j(b_j,h_0)),
}which holds from the additive separability of risk $R_{\infty}$ and $h_j(b_j,0)=h_j(b_j,h_0)=b_j$.

Since we have shown that $\sup_{h} R_{\infty}(\delta_{c^*}, h) = R^*$ for any $h$, we can compute $c_j^*$ by solving $\inf_{c_j} \sup_{b_j} E_{h_j(b_j,0)}L(\delta_{c_j} , h_j(b_j,0))$ for $j=1,\ldots,J$.
\hfill $\square$

\paragraph{Proof of Theorem \ref{thm:asymp_opt_parametric}:}
The proof is composed of multiple steps.

\noindent\textbf{Step 1: }Let $\delta^R$ be a $J\times 1$ vector of decision rules whose $j$-th element is defined $\delta^R_j:= \mathbbm{1}( (\triangledown_\theta  g_j) \Delta / \sigma_{g_j} > 0)$. We show that $\delta^R$ is the minimax solution of the limit experiment.

From the results of Lemma \ref{thm:slicing}, we can find the minimax rule of the limit experiment by solving the cutoff point $c^*_j$ along a slice of $h_j(b_j,0)$ for $j=1,\ldots,J$. Recall that the asymptotic risk function is a linear combination of $R_{j,\infty}$. Thus, we will focus on the following optimization problem:
\eqs{
  \inf_{c_j} \sup_h  R_{j,\infty}(\delta_{j,c_j}^R, h).
}Let $b_j = (\triangledown_\theta  g_j) \Delta / \sigma_{g_j}$. Then, Lemma 5 in \citet{hirano2009asymptotics} implies that the unique solution $c^*_j$ to the optimization problem satisfies
\eqs{
  \sup_{b_j\le0} (-b_j\Phi(b_j-c^*_j)) = \sup_{b_j >0} b_j \Phi(c^*_j -b_j),
}where $\Phi(\cdot)$ is the cdf of the standard normal distribution.
Since both sides have the symmetric structure, we can conclude that $c^*_j=0$.

\noindent\textbf{Step 2: }For any sequence of rules $\delta_{n}$ and the matching rule $\delta$, we show that
\eqs{
  \lim_{n \to \infty} \sqrt{n} R(\delta_{n}, \theta_0 + h/\sqrt{n}) = R_{\infty} (\delta,h).
}
Recall that
\eqs{
  R_{\infty} (\delta,h) := \sum_{j=1}^J R_{j,\infty}(\delta_j,h),
}where
\eqs{
  R_{j,\infty}(\delta_j,h)  & := \int L_{j,\infty} (\delta_j(\Delta),h) dN(\Delta|h, I_0^{-1}), \\
  L_{j,\infty} (\delta_j,h) & := (\triangledown_\theta  g_j) h \left[ \mathbbm{1}\left((\triangledown_\theta  g_j) h>0\right) - \delta_j  \right].
}Note that
\eqs{
  \lim_{n \to \infty} \sqrt{n}R_j(\delta_{j,n},\theta_0+h/\sqrt{n})
  & = \lim_{n \to \infty} \int \sqrt{n}L_{j}(\delta_{j,n}(\omega^n),\theta_0+h/\sqrt{n}) dQ^n_{\theta_0+h/\sqrt{n}} \\
  & = \int (\triangledown_\theta  g_j) h \left[ \mathbbm{1}\left((\triangledown_\theta  g_j) h>0\right) - \delta_j(\Delta)  \right] dN(\Delta|h, I_0^{-1}) \\
  & = R_{j,\infty}(\delta_{j},h).
}Then, the claim is established by the fact that both risk functions $R$ and $R_{\infty}$ are additively separable.

\noindent\textbf{Step 3: }We show that $\delta^R_n$ is matched by $\delta^R$ in the limit experiment.

We can utilize the additive separability property again. Thus, it is enough to show that $\delta^R_{n,j}$ is matched by $\delta^R_{j}$ in the limit experiment. Recall that
\eqs{
  \delta^R_{n,j} =  \mathbbm{1}\lt( \sqrt{n} \frac{ g_j(\hat{\tht}_{n}) }{\hat{\sigma}_{g_j}} > 0\rt)
  ~\mbox{ and }~
  \delta^R_j     =  \mathbbm{1}\lt( \frac{(\triangledown_\theta  g_j) \Delta}{\sigma_{g_j}} > 0\rt).
}Let $S_n$ be a sequence of random variables such that $S_n \stackrel{\theta_0}{\rightsquigarrow} N(0,I_0)$. Since $\hat{\theta}_n$ is best regular and $Q_{\theta}$ is differentiable in quadratic mean, we have
\eqs{
  & \sqrt{n}(\hat{\theta}_n-\theta_0) = I_0^{-1}S_n -\frac{1}{2}h'I_0h + o_{Q_{\theta_0}}(1), \\
  & \log \frac{dQ^n_{\tht_0+h_n/\sqrt{n}}}{dQ^n_{\tht_0}} = h'S_n + \frac{1}{2}h'I_i h + o_{Q_{\theta_0}}(1).
}for $h_n\to h$. Expanding $g_j(\hat{\tht})$ around $\tht_0$ and applying Slutsky's theorem and the delta method, we have
\eqs{
  \lt(\sqrt{n} \frac{ g_j(\hat{\tht}_{n}),  }{\hat{\sigma}_{g_j}}, \log \frac{dQ^n_{\tht_0+h_n/\sqrt{n}}}{dQ^n_{\tht_0}} \rt)
  \stackrel{\theta_0}{\rightsquigarrow}
  N\lt(
  \begin{pmatrix}
  0 \\
  -\frac{1}{2}h'I_0 h
  \end{pmatrix},
  \begin{pmatrix}
  1 & \frac{(\triangledown_\theta  g_j)'h}{\sigma_{g_j}} \\
  \frac{(\triangledown_\theta  g_j)'h}{\sigma_{g_j}} & h'I_0 h
  \end{pmatrix}
  \rt).
}Applying Le Cam's third lemma, we conclude that
\eqs{
  \sqrt{n} \frac{ g_j(\hat{\tht}_{n}),  }{\hat{\sigma}_{g_j}}
  \stackrel{h}{\rightsquigarrow}
  N\lt(
  \frac{(\triangledown_\theta  g_j)'h}{\sigma_{g_j}},
  1
  \rt),
}which establishes the claim.

Finally, the theorem is established by applying Lemma 4 in \citet{hirano2009asymptotics} of which requirements are shown in Steps 2--3 above.
\hfill $\square$

\bibliographystyle{chicago}
\bibliography{reference}

\end{document}